\pdfoutput=1
\documentclass[11pt]{article}
\usepackage{jheppub}
\usepackage{graphicx}
\usepackage{amssymb}
\usepackage{amsmath,amssymb}
\usepackage{slashed}
\usepackage{hyperref}
\usepackage{caption}
\usepackage{xcolor}
\usepackage{dsfont}
\usepackage{verbatim}
\usepackage{subfig}
\usepackage[normalem]{ulem}
\usepackage{amsmath,amssymb,amscd,amsfonts,mathtools}
\usepackage{xcolor}
\definecolor{markgreen}{RGB}{230,243,230}
\definecolor{darkolivegreen}{rgb}{0.33, 0.42, 0.18}
\definecolor{darkpastelgreen}{rgb}{0.01, 0.75, 0.24}

\usepackage[toc,page]{appendix}
\usepackage{epsfig}
\usepackage{epstopdf}
\usepackage{latexsym}
\usepackage{graphicx}
\usepackage{booktabs}
\usepackage{bbm}
\usepackage{color}
\usepackage{physics}
\usepackage{tensor}
\usepackage{verbatim}
\usepackage{subfig}
\usepackage{tikz}
\usepackage{ifthen}
\usetikzlibrary{matrix}
\usetikzlibrary{decorations.markings,calc,shapes,decorations.pathmorphing,patterns,decorations.pathreplacing,arrows.meta}
\usetikzlibrary{positioning}
\usepackage{hyperref}
\usetikzlibrary{patterns}
\usetikzlibrary{positioning}

\makeatletter
\def\@fpheader{\relax}
\makeatother

\newboolean{showcomments}
\setboolean{showcomments}{false}
\newcommand\rem[1]{\ifthenelse{\boolean{showcomments}}{{#1}}{}}
\newcommand{\be}{\begin{equation}}
\newcommand{\ee}{\end{equation}}

\author{Hao Geng}
\affiliation{\textit{Center for the Fundamental Laws of Nature, Harvard University, 17 Oxford St., Cambridge, MA, 02139, USA.}}
\emailAdd{haogeng@fas.harvard.edu}

\title{\Large Open AdS/CFT via a Double-trace Deformation}
\preprint{\today}
\abstract{A concrete model of extracting the physics from the bulk of a gravitational universe is important to the study of quantum gravity and its possible relationship with experiments. Such a model can be constructed in the AdS/CFT correspondence by gluing a bath on the asymptotic boundary of the bulk anti-de Sitter (AdS) spacetime. This bath models a laboratory and is described by a quantum field theory. In the dual conformal field theory (CFT) description this coupling is achieved by a double-trace deformation that couples the CFT with the bath. This suggests that the physics observed by the laboratory is fully unitary. In this paper, we analyze the quantum aspects of this model in detail which conveys new lessons about the AdS/CFT correspondence, and we discuss the potential usefulness of this model in understanding subregion physics in a gravitational universe.}
\begin{document}	
\maketitle
\flushbottom
\pagebreak
\section{Introduction}
Ever since the discovery of the AdS/CFT correspondence \cite{Maldacena:1997re,Witten:1998qj,Gubser:1998bc}, it is hoped that we could learn deep lessons about quantum gravity from it. However, it turns out that the main difficulty is describing a bulk observer and its experience in a reasonable way in the AdS/CFT framework. This question is important as the observer is modeling an experimentalist whose experience is of direct experimental relevance to quantum gravity. However, the description of the observer cannot be so complicated as otherwise it will significantly backreact on the bulk geometry due to the gravitational effect, and meanwhile it cannot be so simple as otherwise, the observer cannot do much to be of practical relevance (see \cite{Almheiri:2012rt,Mathur:2010kx,Mathur:2011wg,Papadodimas:2012aq,Jafferis:2020ora,deBoer:2022zps,Witten:2021unn,Witten:2023qsv} for some attempts to address this question). Besides this subtlety of properly defining the observer with a decent level of complexity, a more basic issue in a gravitational theory is how to define a bulk observer diffeomorphism invariantly. These complications are not anything new, and they also appear in gauge theories, for example quantum electrodynamics.\footnote{Though some properties of gauge theories are encoded in extended gauge invariant operators like Wilson lines.} The usual lore is that the experimentally relevant data are the scattering amplitudes that can be extracted in the asymptotic regime, where the gauge interaction is weak, by putting a detector there. Hence, in quantum gravity, we expect the same, i.e., we should think of the observer as being located at the asymptotic boundary and extracting the relevant data from the bulk to reconstruct the bulk physics using these data. In this paper we will study an explicit model of such an observer constructed using the AdS/CFT correspondence.

\begin{figure}
    \centering
    \begin{tikzpicture}
      \draw[-,very thick,red](0,-2) to (0,2);
       \draw[fill=green, draw=none, fill opacity = 0.1] (0,-2) to (4,-2) to (4,2) to (0,2);
           \draw[-,very thick,red](0,-2) to (0,2);
       \draw[fill=blue, draw=none, fill opacity = 0.1] (0,-2) to (-4,-2) to (-4,2) to (0,2);
       \node at (-2,0)
       {\textcolor{black}{$AdS_{d+1}$}};
        \node at (2,0)
       {\textcolor{black}{$Mink_{d+1}$}};
       \draw [-{Computer Modern Rightarrow[scale=1.25]},thick,decorate,decoration=snake] (-1,-1) -- (1,-1);
       \draw [-{Computer Modern Rightarrow[scale=1.25]},thick,decorate,decoration=snake] (1,1) -- (-1,1);
    \end{tikzpicture}
    \caption{We couple the gravitational AdS$_{d+1}$ universe (the blue shaded region) with a nongravitational bath (the green shaded region) by gluing them along the asymptotic boundary the AdS$_{d+1}$ (the red vertical line). The nongravitational bath is modeled by a $(d+1)$-dimensional CFT living on a half Minkowski space which shares the same boundary as the asymptotic boundary of the AdS$_{d+1}$. We use the Poincar\'{e} coordinates in the AdS$_{d+1}$. The coupling is achieved in the dual CFT description by Equ.~(\ref{eq:nondoublet}).}
    \label{pic:nongravbath}
\end{figure}
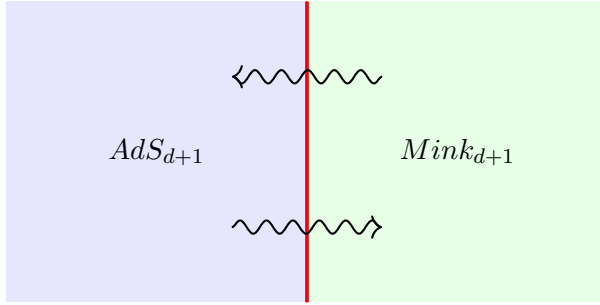

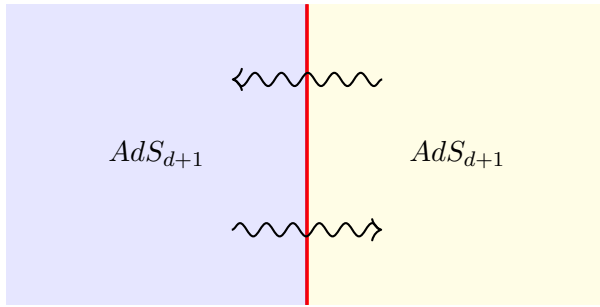
\begin{figure}
    \centering
    \begin{tikzpicture}
       \draw[-,very thick,red](0,-2) to (0,2);
       \draw[fill=yellow, draw=none, fill opacity = 0.1] (0,-2) to (4,-2) to (4,2) to (0,2);
           \draw[-,very thick,red](0,-2) to (0,2);
       \draw[fill=blue, draw=none, fill opacity = 0.1] (0,-2) to (-4,-2) to (-4,2) to (0,2);
       \node at (-2,0)
       {\textcolor{black}{$AdS_{d+1}$}};
        \node at (2,0)
       {\textcolor{black}{$AdS_{d+1}$}};
       \draw [-{Computer Modern Rightarrow[scale=1.25]},thick,decorate,decoration=snake] (-1,-1) -- (1,-1);
       \draw [-{Computer Modern Rightarrow[scale=1.25]},thick,decorate,decoration=snake] (1,1) -- (-1,1);
    \end{tikzpicture}
    \caption{We couple the gravitational AdS$_{d+1}$ universe (the blue shaded region) with a gravitational bath (the yellow shaded region) by gluing them along the asymptotic boundary the AdS$_{d+1}$ (the red vertical line). The gravitational bath is modeled by another  AdS$_{d+1}$ which shares the same asymptotic boundary with the original AdS$_{d+1}$. We take the Poincar\'{e} coordinates in both of the AdS$_{d+1}$. The coupling is achieved in the dual CFT description by Equ.~(\ref{eq:gravdoubet}).}
    \label{pic:gravbath}
\end{figure}

More specifically, we consider a gravitational theory in an asymptotically AdS$_{d+1}$ spacetime, which has a dual CFT$_{d}$ description that resides on the asymptotic boundary of the AdS$_{d+1}$. We model the observer by a $(d+1)$-dimensional bath that is glued to the asymptotic boundary of the AdS$_{d+1}$ (see \cite{Rocha:2008fe,Jana:2020vyx,Loganayagam:2022zmq,Shen:2023srk,Pelliconi:2023ojb,Karch:2023wui,Kehrein:2023yeu} for early and recent studies on some aspects of relevant models). This bath can be either gravitational or non-gravitational. When it is non-gravitational, we model it by a $(d+1)$-dimensional CFT living on a half Minkowski space whose boundary is glued with the asymptotic boundary of the AdS$_{d+1}$ (see Fig.~\ref{pic:nongravbath}). This gluing can be more easily described using the dual CFT$_{d}$ where we put the CFT$_{d}$ on the boundary of the bath CFT$_{d+1}$ and couple them by a double-trace deformation
\begin{equation}
S_{tot}=S_{\text{CFT$_{d}$}}+S_{\text{CFT$_{d+1}$}}+h\int d^{d}x \mathcal{O}_{1}(t,\vec{x})\mathcal{O}_{2}(t,\vec{x})\,,\label{eq:nondoublet}
\end{equation}
where $O_{1}(t,\vec{x})$ is a CFT$_{d}$ single-trace scalar primary operator and $O_{2}(t,\vec{x})$ is a CFT$_{d+1}$ single-trace scalar primary operator extrapolated to the boundary with a coupling constant $h$. We will take the double-trace deformation to be marginal (i.e., the sum of the conformal weights of $\mathcal{O}_{1}(t,\vec{x})$ and $\mathcal{O}_{2}(t,\vec{x})$ satisfies $\Delta_{1}+\Delta_{2}=d$). When the bath is gravitational (see Fig.\ref{pic:gravbath}), we will take its geometry to be another AdS$_{d+1}$ which also has a dual description as another CFT$_{d}$ which we will call CFT$_{d}^{2}$. Then the gluing of this AdS$_{d+1}$ bath to the original AdS$_{d+1}$ can also be described in the CFT language by a double-trace deformation
\begin{equation}
S_{tot}=S_{\text{CFT$_{d}$}}+S_{\text{CFT$_{d}^{2}$}}+\int d^{d}x\mathcal{O}_{1}(t,\vec{x})\mathcal{O}_{2}(t,\vec{x})\,,\label{eq:gravdoubet}
\end{equation}
where again $\mathcal{O}_{1}(t,\vec{x})$ is a single-trace scalar primary operator of the CFT$_{d}$ and now $\mathcal{O}_{2}(t,\vec{x})$ is a single-trace scalar primary operator of the CFT$_{d}^{2}$ and we take this deformation to be marginal.

We will see that in the dual AdS$_{d+1}$ description, the double-trace deformation is modifying the boundary condition of the free massive scalar field that duals to $\mathcal{O}_{1}(t,\vec{x})$ \cite{Witten:2001ua}. We will show that this modified boundary condition induces several interesting quantum effects in the AdS$_{d+1}$ bulk. The first effect is that the bulk Hilbert space constructed from the AdS$_{d+1}$ scalar field (the Fock space) is twice as large as the case in the standard AdS/CFT. This is easily understood if we notice that the bulk AdS$_{d+1}$ is coupled to a bath, and hence the particles from the bath are free to enter the AdS$_{d+1}$ which, together with the original particles in the AdS$_{d+1}$ double (or square) the Hilbert space. This can be seen if we perform canonical quantization with the modified boundary condition for the AdS$_{d+1}$ scalar field. The second effect is that the modified boundary condition of the bulk scalar field disables us to define a diffeomorphism invariant path integral measure for the scalar field. Hence, there could be a diffeomorphism anomaly that should be compensated if we want to consistently couple the scalar field to gravity. We will show that there is a natural mechanism to compensate this diffeomorphism anomaly. This mechanism turns out to be a St\"{u}ckelberg mechanism to generate a mass for the graviton, and this suggests that the diffeomorphism invariance is spontaneously broken due to the bath coupling. This is a rather remarkable result as it shows that the gravitational theory is modified due to the existence of the observer or because the gravitational universe is being observed. To uncover all these effects, we start with clarifications of several subtleties in the AdS/CFT correspondence.

Moreover, in the gravitational bath case the original AdS$_{d+1}$ can be thought of a subregion of a gravitational universe (which is the union of the original AdS$_{d+1}$ and the bath AdS$_{d+1}$).\footnote{This perspective was adopted at a classical level in \cite{Compere:2019bua,Compere:2020lrt,Fiorucci:2020xto} by considering leaky boundary conditions instead of the standard reflective ones.} Therefore, the analysis presented here is potentially helpful in understanding subregion physics in a gravitational universe.\footnote{See \cite{Donnelly:2016auv,Speranza:2017gxd,Kirklin:2018gcl,Camps:2018wjf,Camps:2019opl,Bahiru:2022oas,Bahiru:2023zlc,Jensen:2023yxy,Folkestad:2023cze,balasubramanian2023entropy} for early and recent attempts in setting up and understanding this difficult question.}

This paper is organized as follows. In Sec.~\ref{sec:review} we review some basics of the AdS/CFT correspondence such as the calculation of the boundary CFT correlator using the AdS bulk and we clarify some subtleties that are relevant to the quantum aspects of the bulk scalar field-- the canonical quantization and the path integral quantization. In Sec.~\ref{sec:dbt} we review and refine the double-trace deformation in AdS/CFT \cite{Witten:2001ua,Aharony:2001pa,Berkooz:2002ug} by providing a path integral derivation of the proposal in \cite{Witten:2001ua} and check the consistency with the results in Sec.~\ref{sec:review}. In Sec.~\ref{sec:dbto} we use the techniques we developed in Sec.~\ref{sec:review} and Sec.~\ref{sec:dbt} to analyze the two aforementioned models of coupling AdS$_{d+1}$ to a bath and we show the first quantum effect-- the Hilbert space doubling. In Sec.~\ref{sec:PIDI} we discuss the second quantum effect--the issue of diffeomorphism invariance broken and restoration. In Sec.~\ref{sec:subr} we discuss how the model we are considering in this paper could be potentially useful to the study of subregion physics in a gravitational universe. In Sec.~\ref{sec:conclusion} we conclude the paper with a summary. In Appendix.\ref{sec:A} we discuss a potential loophole in our consideration in Sec.~\ref{sec:PIDI} and its resolution.

\section{Quantization Schemes and Propagating Modes in AdS/CFT}\label{sec:review}
In this section, we review the computation of the generating functional for correlators of the dual single-trace scalar primary operator in the CFT$_{d}$ from the massive scalar field in the AdS$_{d+1}$. Then we analyze the bulk propagating modes of the scalar field carefully distinguishing the on-shell and off-shell modes.  For simplicity we consider a free minimally coupled massive scalar field $\phi(x)$ and our treatment will be in the framework of \cite{Berkooz:2002ug}.

We take the geometry of AdS$_{d+1}$ to be in the Poincar\'e patch of the Lorentzian signature:
\begin{equation}
ds^2=g_{\mu\nu}dx^{\mu}dx^{\nu}=\frac{ z^2+\sum_{i=1}^{d}\eta_{ab}dx^{a}dx^{b}}{z^2}\label{eq:metric}\,,
\end{equation}
and the action for the scalar field $\phi(x)$ is 
\begin{equation}
S_{0}=\frac{1}{2}\int \sqrt{-g}dzdtd^{d-1}\vec{x}\phi(x)\Big(\Box-m^2\Big)\phi(x)\label{eq:action}\,,
\end{equation}
where $\Box=\frac{1}{\sqrt{-g}}\partial_{\mu}(g^{\mu\nu}\sqrt{-g}\partial_{\nu})$.\footnote{For simplicity of the analysis, we don't consider self-interactions of the scalar field in this paper. Self-interactions can be treated perturbatively once we have had a good understanding of the free limit.} The asymptotic boundary of AdS$_{d+1}$, where the dual CFT$_{d}$ lives, is at $z\rightarrow0$ and we will take it to be at $z=\epsilon$ and send $\epsilon\rightarrow0$ at the end of our calculation. We emphasize on the calculational details and when we are able to perform the alternative quantization.

According to the AdS/CFT correspondence \cite{Maldacena:1997re,Gubser:1998bc,Witten:1998qj}, the scalar field $\phi(x)=\phi(z,t,\vec{x})$ duals to a single-trace scalar primary operator $\mathcal{O}(t,\vec{x})$ in the CFT$_d$ and imposing a boundary condition $J(t,\vec{x})$ for the scalar field $\phi(x)$ is equivalent to turning on a source for $\mathcal{O}(t,\vec{x})$ (see later discussions for details). Moreover, this duality states that
\begin{equation}
 Z[J]_{\text{CFT}}\equiv\langle e^{i\int dtd^{d-1}\vec{x} J(t,\vec{x})\mathcal{O}(t,\vec{x})}\rangle_{\text{CFT}}=Z[J]_{AdS_{d+1}}\,,
\end{equation}
where $Z[J]_{\text{CFT}}$ is the generating functional for correlators of the operator $\mathcal{O}(t,\vec{x})$ on the CFT$_{d}$ side and $Z[J]_{AdS_{d+1}}$ is the partition function of the full gravitational theory (string theory) in AdS$_{d+1}$ with the specified boundary condition $J(t,\vec{x})$ for the scalar field $\phi(x)$. For our purpose (in the low energy regime and with a weak gravitational coupling), $Z[J]_{AdS_{d+1}}$ can be approximated by the partition function of the scalar field $\phi(x)$ in the fixed AdS$_{d+1}$ background
\begin{equation}
 Z[J]_{AdS_{d+1}}=\int D[\phi;J]e^{-iS_{J}[\phi]}\,,
\end{equation}
where we restrict ourselves to consider scalar field configurations satisfying the boundary condition $J(t,\vec{x})$ and $S_{J}[\phi]$ is given by $S_{0}$ (in Equ.~(\ref{eq:action})) plus appropriate boundary terms. The boundary terms will be $J(t,\vec{x})$ dependent. Since we are considering a free field $\phi(x)$ the calculation reduces to the computation of the on-shell action $S^{\text{on-shell}}[\phi]$ with boundary condition $J(t,\vec{x})$
\begin{equation}
   Z[J]_{\text{CFT}}\equiv\langle e^{i\int dtd^{d-1}\vec{x} J(t,\vec{x})\mathcal{O}(t,\vec{x})}\rangle_{\text{CFT}}=Z[J]_{AdS_{d+1}}\sim e^{-iS^{\text{on-shell}}_{J}[\phi_{J}]}\,,\label{eq:gf}
\end{equation}
where $\phi_{J}(x)$ satisfies the boundary condition $J(t,\vec{x})$ (to be specified later) and the equation of motion
\begin{equation}
 (\Box-m^2)\phi_{J}(x)=0\,\label{eq:eom}.
\end{equation}
\subsection{The Meaning of Quantization}\label{sec:meanq}

Before we discuss different quantization schemes in AdS/CFT in this section, let's articulate the meaning of quantization. In the standard treatment of a quantum system, quantization means constructing the Hilbert space of the system or specifying the rule to calculate all the transition amplitudes. 

The former is usually done for a free theory using the canonical quantization for which in our case we have to solve the equation Equ.~(\ref{eq:action}) with vanishing $J(t,\vec{x})$, find all the solutions that satisfy the boundary condition $J(t,\vec{x})=0$, and we will call these solutions as on-shell modes, then assign creation and annihilation operators for each of the modes and their complex conjugates. Schematically, let's denote these modes in the Fourier space by $\delta\phi_{n}(z,\omega_{n,\vec{k}},\vec{k})$ and the field operator $\hat{\phi}(z,t,\vec{x})$ is given by
\begin{equation}
\hat{\phi}(z,t,\vec{x})=\int \frac{d\vec{k}}{(2\pi)^{d-1}}\sum_{n}\frac{1}{\sqrt{2\omega_{n,\vec{k}}}}\Big(\delta\phi(z,\omega_{n,\vec{k}},\vec{k})\hat{a}_{n\vec{k}}e^{i\omega_{n,\vec{k}}t-i\vec{k}\cdot\vec{x}}+\delta\phi^{*}(z,\omega_{n,\vec{k}},\vec{k})\hat{a}^{\dagger}_{n\vec{k}}e^{-i\omega_{n,\vec{k}}t+i\vec{k}\cdot\vec{x}}\Big)\,.\label{eq:canonicalq}
\end{equation}
Then imposing the canonical quantization condition
\begin{equation}
[\hat{\phi}(z,t,\vec{x}),g^{tt}\partial_{t}\hat{\phi}(z',t,\vec{x'})]=\frac{i}{\sqrt{-g}}\delta(z-z')\delta^{d-1}(\vec{x}-\vec{x'})\,,\label{eq:canonicalqu}
\end{equation}
we will get the standard commutation relations for the creation and annihilation operators
\begin{equation}
[\hat{a}_{n,\vec{k}},\hat a_{n',\vec{k'}}]=0\,,\quad [\hat{a}_{n,\vec{k}},\hat a^{\dagger}_{n',\vec{k'}}]=\delta_{nn'}\delta^{d-1}(\vec{k}-\vec{k'})\,,
\end{equation}
given that the on-shell modes are orthonormal in the so-called Klein-Gordon measure
\begin{equation}
(\delta\phi,\delta\phi)_{\text{KG}}=i\int_{\Sigma} dz d^{d-1}x\sqrt{-g}g^{tt}(\delta\phi^{*}\partial_{t}\delta\phi-\delta\phi\partial_{t}\delta\phi^{*})(z,t,\vec{x})\,.\label{eq:onmeasure}
\end{equation}
In the end, the Hilbert space is constructed as the Fock space of these annihilation and creation operators. 

The later is asking for a calculational scheme for correlators of the theory which is usually done using the path integral formalism. The correlators are easily generated once we know the generating functional Equ.~(\ref{eq:gf}). In our case, we have to compute the path integral of a free scalar field in the AdS$_{d+1}$ with a specified boundary condition $J(t,\vec{x})$. Since we are considering a free scalar field, the $J(t,\vec{x})$-dependent part of this calculation reduces to the computation of the on-shell action $S_{J}^{\text{on-shell}}[\phi_{J}]$ with the boundary condition $J(t,\vec{x})$. Nevertheless, the full partition function contains a one-loop correction which though doesn't contribute to the generating functional as it is $J(t,\vec{x})$-independent. Since we will study quantum effects later, we keep track of everything. We have
\begin{equation}
\begin{split}
    Z[J]_{\text{AdS}_{d+1}}&=\int D[\phi;J]e^{-iS_{J}[\phi]}\\&=\int D[\delta\phi]e^{-i\frac{1}{2}\int\sqrt{-g}d^{d+1}x \delta\phi(\Box-m^2)\delta\phi}e^{-iS^{\text{on-shell}}_{J}[\phi_{J}]}\\&=\text{det}^{-\frac{1}{2}}(\Box-m^2)e^{-iS^{\text{on-shell}}_{J}[\phi_{J}]}\,,\label{eq:quantumZ}
    \end{split}
\end{equation}
where the $\delta\phi$ describes the fluctuations above the background on-shell solution and $\delta\phi$ satisfies a boundary condition with vanishing $J(t,\vec{x})$ (the precise boundary condition will be discussed later). The determinant is a symbolic representation of the path integral over the fluctuations. This path integral can be performed as follows. Firstly, we solve for the eigenmodes ($\delta\phi_{\lambda}(z,\omega_{\lambda,n,\vec{k}},\vec{k})e^{-i\omega_{\lambda,n,\vec{k}}t-i\vec{k}\cdot \vec{x}}$) associated with all the nonzero eigenvalues $\lambda$ (in fact we need $\lambda>0$, see the next subsections) of the operator $\Box-m^2$
\begin{equation}
(\Box-m^2)\delta\phi_{\lambda}(z,\omega_{\lambda,n,\vec{k}},\vec{k})e^{-i\omega_{\lambda,n,\vec{k}}t+i\vec{k}\cdot \vec{x}}=\lambda\delta\phi_{\lambda}(z,\omega_{\lambda,n,\vec{k}},\vec{k})e^{-i\omega_{\lambda,n,\vec{k}}t+i\vec{k}\cdot \vec{x}}.\label{eq:modeeom}
\end{equation}
These modes are orthonormal for both $n$ and $\lambda$ 
\begin{equation}
\int \sqrt{-g}d^{d+1}x \delta\phi^{*}_{\lambda}(z,\omega_{\lambda,n,\vec{k}},\vec{k})\delta\phi_{\lambda'}(z,\omega_{\lambda',n',\vec{k'}})e^{-i(\omega_{\lambda',n',\vec{k'}}-\omega_{\lambda,n,\vec{k}})t+i(\vec{k'}-\vec{k})\cdot\vec{x}}=\delta_{\lambda\lambda'}\delta_{nn'}\delta^{d-1}(\vec{k}-\vec{k'}).\label{eq:offmeasure}
\end{equation}
As we will discuss in detail in the next section, the normalizability of these modes is important for determining the proper boundary conditions or the fall-off behaviors for them when $z\rightarrow0$. Then we use these modes to expand a generic function $\delta\phi(z,t,\vec{x})$ with the appropriate boundary condition as
\begin{multline}
    \delta\phi(z,t,\vec{x})=\int \frac{d\vec{k}}{(2\pi)^{d-1}}a_{0,n,\vec{k}}\delta\phi_{\lambda}(z,\omega_{n,\vec{k}},\vec{k})e^{-i\omega_{\lambda,n,\vec{k}}t+i\vec{k}\cdot \vec{x}} \\
    +\int \frac{d\vec{k}}{(2\pi)^{d-1}}\sum_{\lambda,n} a_{\lambda,n,\vec{k}}\delta\phi_{\lambda}(z,\omega_{\lambda,n,\vec{k}},\vec{k})e^{-i\omega_{\lambda,n,\vec{k}}t+i\vec{k}\cdot \vec{x}}\,,
\end{multline}
where $a_{\lambda,n,\vec{k}}$ are complex numbers that ensure the reality of $\delta\phi(z,t,\vec{x})$. Now we can compute
\begin{equation}
\begin{split}
\int D\delta\phi e^{-i\frac{1}{2}\int\sqrt{-g}d^{d+1}x\delta\phi(\Box-m^2)\delta\phi}=&\int \Pi_{\lambda,n,\vec{k}}da_{\lambda,n,\vec{k}}e^{-i\frac{1}{2}\sum_{\lambda,n,\vec{k}}|a_{\lambda,n,\vec{k}}|^2\lambda}\\=&\Pi_{\lambda,n,\vec{k}}(\frac{1}{\sqrt{i\lambda}})\\=&\text{det}^{-\frac{1}{2}}(\Box-m^2)\,,\label{eq:1loop}
\end{split}
\end{equation}
where for the simplicity of notation we have discretized the $\vec{k}$ by putting the $\vec{x}$ in a box. The important step is the second step for which we transform the functional integration measure $D\delta\phi$ to the integral of the coefficients $\Pi_{\lambda,n,\vec{k}}da_{\lambda,n\vec{k}}$. This step in general requires a Jacobian which is an overall constant (or an ambiguity in the definition of a functional integration measure) and is fixed by 
\begin{equation}
\int D\delta\phi e^{-\int \sqrt{-g}d^{d+1}x\delta\phi^2}=1\,.\label{eq:measure}
\end{equation}
Here we notice a subtlety that the on-shell modes part $a_{0,n,\vec{k}}$ don't contribute to the integrand in Equ.~(\ref{eq:1loop}) but they do contribute to the functional integral Equ.~(\ref{eq:measure}) in the definition of the path integral measure when the boundary source is zero. This subtlety is important for our later study in Sec.~\ref{sec:PIDI} and we will ponder it till Sec.~\ref{sec:PIDI}.

\subsection{Standard Quantization in AdS/CFT}
In this subsection, we review the standard quantization scheme for a massive free scalar field in AdS$_{d+1}$. Such a scalar field is dual to a single-trace scalar primary operator $O(t,\vec{x})$ in the boundary CFT$_d$. We review the computation of generating functional for this single-trace operator using the AdS$_{d+1}$ description Equ.~(\ref{eq:gf}) and the behavior of the on-shell and off-shell modes near $z\rightarrow0$ which are relevant respectively to the construction of the Hilbert space and the quantum correction to the bulk partition function.
\subsubsection{Partition Function}\label{sec:partitons}
The equation of motion Equ.~(\ref{eq:eom}) comes from setting the variation of the action Equ.~(\ref{eq:action}) to zero. However, there are additional boundary terms in the variation (suppose that the equation of motion has been satisfied)
\begin{equation}
\delta S_{0}=-\frac{1}{2}\int_{z=\epsilon}d^{d}\vec{x}z^{-d+1}\Big(\delta\phi\partial_{z}\phi-\phi\partial_{z}\delta\phi\Big)\,.
\end{equation}
Adding a specific boundary term $S_{bdy}$ to the action $S_{0}$ and setting $\delta S_{0}+\delta S_{bdy}=0$ specify the boundary condition for the scalar field $\phi(x)$. For the standard quantization we will choose 
\begin{equation}
  S_{bdy}=-\epsilon^{-\Delta_{+}}\pi^{\frac{d}{2}}\frac{\Gamma[\Delta_{+}+1-\frac{d}{2}]}{\Gamma[\Delta_{+}]}\int dtd^{d-1}\vec{x}\phi(\epsilon,t,\vec{x})J(t,\vec{x})\,,
\end{equation}
which gives the boundary condition $J(t,\vec{x})$:
\begin{equation}
  z^{-\Delta_{-}}(z\partial_{z}-\Delta_{+})\phi(x)|_{z=\epsilon\rightarrow0}=-2\pi^{\frac{d}{2}}\frac{\Gamma[\Delta_{+}+1-\frac{d}{2}]}{\Gamma[\Delta_{+}]}J(t,\vec{x})\,,\label{eq:bcs}
\end{equation}
for a fixed $J(t,\vec{x})$ and a generic fluctuation mode should satisfy
\begin{equation}
 z^{-\Delta_{-}}(z\partial_{z}-\Delta_{+})\delta\phi(x)=0\,,\quad\text{as $z\rightarrow0$}\,,\label{eq:fbcs}
\end{equation}
where we have defined $\Delta_{\pm}=\frac{d}{2}\pm\sqrt{\frac{d^2}{4}+m^2}$. 

A solution of the equation of motion $\phi_{J}(x)$ which is regular in interior of the AdS$_{d+1}$ \cite{Witten:1998qj} and satisfies the boundary condition Equ.~(\ref{eq:bcs}) is
\begin{equation}
\phi_{J}(x)=\int dt'd^{d-1}\vec{x'}\frac{J(t',\vec{x'})z^{\Delta_{+}}}{(-(t-t')^2+z^2+|\vec{x}-\vec{x'}|^2)^{\Delta_{+}}}\,,\label{eq:solutions}
\end{equation}
which is uniquely specified.\footnote{This explains the statement in Sec.~\ref{sec:meanq} that the boundary condition $J(t,\vec{x})$ fixed the on-shell profile and hence the on-shell modes don't contribute to the path integral.} To check that Equ.~(\ref{eq:solutions}) satisfies the boundary condition Equ.~(\ref{eq:bcs}) it is important to know that
\begin{equation}
 \lim_{z\rightarrow0}\frac{z^{2\Delta-d}}{(z^2-(t-t')^{2}+|\vec{x}-\vec{x'}|^2)^{\Delta}}=\pi^{\frac{d}{2}}\frac{\Gamma[\Delta-\frac{d}{2}]}{\Gamma[\Delta]}\delta(t-t')\delta^{(d-1)}(\vec{x}-\vec{x'})\,,\quad\text{for }\quad\Delta>\frac{d}{2}\,.
\end{equation}
Now we can evaluate the on-shell action 
\begin{equation}
\begin{split}
  S_{J}^{\text{on-shell}}[\phi_{J}] &=S_{0}[\phi_{J}]+S_{bdy}[\phi_{J}]  \\
  &=-\pi^{\frac{d}{2}}\frac{\Gamma[\Delta_{+}+1-\frac{d}{2}]}{\Gamma[\Delta_{+}]}\int dtd^{d-1}\vec{x}dt'd^{d-1}\vec{x'}\frac{J(t',\vec{x'})J(t,\vec{x})}{(-(t-t')^2+|\vec{x}-\vec{x'}|^{2})^{\Delta_{+}}}\,,
\end{split}
\end{equation}
and then we get the expected generating functional from Equ.~(\ref{eq:gf})
\begin{equation}
  Z[J]_{\text{CFT}}=e^{i\pi^{\frac{d}{2}}\frac{\Gamma[\Delta_{+}+1-\frac{d}{2}]}{\Gamma[\Delta_{+}]}\int dtd^{d-1}\vec{x}dt'd^{d-1}\vec{x'}\frac{J(t',\vec{x'})J(t,\vec{x})}{(-(t-t')^2+|\vec{x}-\vec{x'}|^{2})^{\Delta_{+}}}}\,,\label{eq:treeZs}
\end{equation}
which is indeed the CFT generating functional for a conformal primary operator of conformal weight $\Delta_{+}$. 
\subsubsection{On-Shell and Off-Shell Modes}\label{sec:prop}
We learned in the previous discussion that the on-shell modes are relevant to the canonical quantization and the off-shell modes are relevant to the calculation of the quantum corrections to the partition function. Moreover, they are normalizable under two different measures Equ.~(\ref{eq:onmeasure}) and Equ.~(\ref{eq:offmeasure}).  Here we will give a detailed study of their properties near the asymptotic boundary $z\rightarrow0$. We study the on-shell modes first and then the off-shell modes.

To understand the on-shell modes 
we consider a solution of the scalar field equation of motion 
\begin{equation}
 (\Box-m^2)\delta\phi(x)=0\,.\label{eq:eomf}
\end{equation}
In the Lorentzian signature, a general solution of Equ.~(\ref{eq:eomf}) is given by \footnote{Here we remember that there is a subtlety as emphasized in \cite{Balasubramanian:1998sn} that the solutions displayed in Equ.~(\ref{eq:sfl}) are those for $\sqrt{\frac{d^2}{4}+m^2}$ nonintegral. We will ignore this subtlety as we only care about the asymptotic behavior as $z\rightarrow0$.} 
\begin{equation}
 \delta\phi(x)_{\omega,\vec{k}}^{\pm}=e^{-i\omega t+i \vec{k}\cdot\vec{x}}z^{\frac{d}{2}}J_{\pm\sqrt{\frac{d^2}{4}+m^2}}(z\sqrt{\omega^2-\vec{k}^2})\,,\label{eq:sfl}
\end{equation}
where $\omega^{2}-\vec{k}^2\geq0$ and we used the Bessel's functions of the first kind $J_{\pm\alpha}$ for $\alpha=\sqrt{\frac{d^2}{4}+m^2}$. We therefore have the asymptotic behavior
\begin{equation}
 \delta\phi(x)_{\omega,\vec{k}}^{\pm}\sim z^{\Delta_{\pm}},\quad\text{as}\quad z\rightarrow0\,.\label{eq:asymos}
\end{equation}

As we discussed in Sec.~\ref{sec:meanq}, the on-shell modes should be normalizable (finite) in the Klein-Gordon measure. The finiteness of the Klein-Gordon measure requires a finite asymptotic behavior
\begin{multline}
(\delta\phi^{\pm},\delta\phi^{\pm})_{\text{KG}}=i\int_{\Sigma} dz d^{d-1}x\sqrt{-g}g^{tt}(\delta\phi^{*\pm}\partial_{t}\delta\phi^{\pm}-\delta\phi^{\pm}\partial_{t}\delta\phi^{*\pm})(z,\vec{x})\\
\sim \int dz z^{-d+1}z^{2\Delta_{\pm}}\,,\quad\text{near}\quad z=0\,,
\end{multline}
where we have used Equ.~(\ref{eq:asymos}). We can see that this finiteness condition requires that $\Delta_{\pm}>\frac{d-2}{2}$ and is trivially satisfied for $\Delta_{+}$. The standard quantization is the scheme that projects out $\delta\phi^{-}$. Hence in this quantization scheme the on-shell modes satisfy the following fall-off behavior
\begin{equation}
\delta\phi_{\text{on-shell}}(x)\sim z^{\Delta_{+}}\,,\quad\text{as $z\rightarrow0$}\,.
\end{equation}

Now let's study the off-shell modes. An off-shell mode satisfies the equation
\begin{equation}
(\Box-m^2)\delta\phi(x)=\lambda\delta\phi(x)\,,\label{eq:offshell1}
\end{equation}
and the normalizability condition
\begin{equation}
||\delta\phi||^2=\int \sqrt{-g}d^{d+1}x |\delta\phi(x)|^2=1\,.\label{eq:offshell2}
\end{equation}
Another condition is that we should have
\begin{equation}
\begin{split}
S_{J}[\phi_{J}+\delta\phi]&=S_{0}[\phi_{J}+\delta\phi]+S_{bdy}[\phi_{J}+\delta\phi]\\&=\frac{1}{2}\int \sqrt{-g}d^{d+1}x\delta\phi(\Box-m^2)\delta\phi+S_{J}^{\text{on-shell}}[\phi_{J}]\,,
\end{split}
\end{equation}
which ensures that Equ.~(\ref{eq:quantumZ}) is consistent.  This requires that Equ.~(\ref{eq:fbcs}) is satisfied and
\begin{equation}
 S_{bdy}[\delta\phi]=-\epsilon^{-\Delta_{+}}\pi^{\frac{d}{2}}\frac{\Gamma[\Delta_{+}+1-\frac{d}{2}]}{\Gamma[\Delta_{+}]}\int dtd^{d-1}\vec{x}\delta\phi(\epsilon,t,\vec{x})J(t,\vec{x})\rightarrow0\,,\label{eq:condition2}
\end{equation}
as $\epsilon\rightarrow0$. Similar to the on-shell case, a general solution of Equ.~(\ref{eq:offshell1}) satisfies 
\begin{equation}
\delta\phi(x)^{\pm}\sim z^{\Delta_{\pm}(\lambda)}\,,\quad\text{as $z\rightarrow0$}\,,
\end{equation}
where $\Delta_{\pm}(\lambda)=\frac{d}{2}\pm\frac{1}{2}\sqrt{d^2+4m^2+4\lambda}$. Then the normalizability condition Equ.~(\ref{eq:offshell2}) requires that $\Delta^{-}(\lambda)$ is projected out for which Equ.~(\ref{eq:fbcs}) is automatically satisfied and Equ.~(\ref{eq:condition2}) sets $\lambda>0$. Hence, in the standard quantization the off-shell modes are constrained to satisfy
\begin{equation}
\delta\phi_{\text{off-shell}}(z)\sim z^{\Delta_{+}(\lambda)}\,,\quad \text{as $z\rightarrow0$}\,,\label{eq:offshellst}
\end{equation}
and $\lambda>0$.

With the knowledge of the on-shell modes, we can construct the Hilbert space by doing the canonical quantization Equ.~(\ref{eq:canonicalqu}). The field operator $\hat{\phi}$(x) is given by
\begin{equation}
\hat{\phi}(x)=\int\frac{d\vec{k}}{(2\pi)^{d-1}}\int_{\omega^2-\vec{k}^2\geq0}\frac{d\omega}{2\pi}\Big(\delta\phi(x)_{\omega,\vec{k}}^{+}\hat{a}^{\dagger}_{\omega,\vec{k}}+\delta\phi(x)^{+*}_{\omega,\vec{k}}\hat{a}_{\omega,\vec{k}}\Big)\,,\label{eq:canonicals}
\end{equation}
where $\hat{a}^{\dagger}_{\omega,\vec{k}}$ and $\hat{a}_{\omega,\vec{k}}$ are the creation and annihilation operators. They satisfy the standard commutation relations
\begin{equation}
[\hat{a}_{\omega,\vec{k}},\hat{a}^{\dagger}_{\omega,\vec{k'}}]=\delta(\omega-\omega')\delta^{d-1}(\vec{k}-\vec{k'})\,.
\end{equation}

With the knowledge of the off-shell modes, we can understand the quantum correction to the partition function Equ.~(\ref{eq:treeZs}). A general field configuration can be expanded as
\begin{equation}
\begin{split}
\phi(x)=&\phi_{J}(x)+\sum_{\lambda>0}\int_{\omega^2-\vec{k}^2\geq0}  \frac{d\vec{k}}{(2\pi)^{d-1}}\int d\omega a_{\lambda,\omega,\vec{k}}\delta\phi_{\lambda}(z,\omega,\vec{k})e^{-i\omega t-i\vec{k}\cdot \vec{x}}\,,
\end{split}
\end{equation}
where we note that we don't have on-shell fluctuation modes as the on-shell profile has been fixed by $\phi_{J}(x)$. The action evaluated on this configuration is
\begin{equation}
\begin{split}
S[\phi] = &S_{0}[\phi]+S_{bdy}[\phi]\\=& -\pi^{\frac{d}{2}}\frac{\Gamma[\Delta_{+}+1-\frac{d}{2}]}{\Gamma[\Delta_{+}]}\int dtd^{d-1}\vec{x}dt'd^{d-1}\vec{x'}\frac{J(t',\vec{x'})J(t,\vec{x})}{(-(t-t')^2+|\vec{x}-\vec{x'}|^{2})^{\Delta_{+}}} \\
&+\frac{1}{2}\sum_{\lambda>0}\int_{\omega^2-\vec{k}^2\geq0} \frac{d^{d-1}\vec{k}}{(2\pi)}d\omega\lambda|a_{\lambda, \omega,\vec{k}}|^2\,.
\end{split}
\end{equation}
The first term gives us the result Equ.~(\ref{eq:treeZs}) and the second term describes the quantum fluctuations. The quantum correction to the partition function is obtained by the Gaussian integral over the coefficients $a_{\lambda,\omega,\vec{k}}$ as in Equ.~(\ref{eq:quantumZ}) and Equ.~(\ref{eq:1loop}).

\subsection{Alternative Quantization in AdS/CFT}

\subsubsection{Review of the Klebanov-Witten Proposal}
Before we carry out a general analysis for the alternative quantization, let's review the Klebanov-Witten proposal for the partition function \cite{Klebanov:1999tb}. Klebanov and Witten suggested that when the mass square of the AdS$_{d+1}$ scalar satisfies the condition that $\Delta_{-}>\frac{d-2}{2}$ (which is the unitarity bound of primary operators in CFT$_d$ \cite{Rychkov:2016iqz,Minwalla:1997ka}) we can perform the so-called alternative quantization such that the dual single-trace scalar primary operator in CFT$_{d}$ has conformal weight $\Delta_{-}$. They also suggested that the generating functional in this case is a Legendre transform of that in the standard quantization. Namely, we have
\begin{equation}
Z[J']_{\text{CFT}}=\int DJ e^{i\pi^{\frac{d}{2}}\frac{\Gamma[\Delta_{+}+1-\frac{d}{2}]}{\Gamma[\Delta_{+}]}\int dtd^{d-1}\vec{x}dt'd^{d-1}\vec{x'}\frac{J(t',\vec{x'})J(t,\vec{x})}{(-(t-t')^2+|\vec{x}-\vec{x'}|^{2})^{\Delta_{+}}}+\pi^{\frac{d}{2}}\frac{\Gamma[\Delta_{+}+1-\frac{d}{2}]}{\Gamma[\Delta_{+}]}\int dtd^{d-1}\vec{x} J(t,\vec{x})J'(t,\vec{x})}\,,
\end{equation}
which can be easily shown, by going to Fourier space, to be
\begin{equation}
 Z[J']_{\text{CFT}}=e^{-i\frac{(2\Delta_{-}-d)\Gamma[\Delta_{-}]}{\pi^{\frac{d}{2}}\Gamma[\Delta_{-}-\frac{d}{2}]}\int dtd^{d-1}\vec{x}dt'd^{d-1}\vec{x'}\frac{J'(t,\vec{x})J'(t',\vec{x'})}{(-(t-t')^2+|\vec{x}-\vec{x'}|^{2})^{\Delta_{-}}}}\,,\label{eq:altgen}
\end{equation}
as a legitimate CFT$_{d}$ generating functional for correlators of a scalar primary operator of conformal weight $\Delta_{-}$.

This calculation can be nicely interpreted in the following way. We first perform the standard quantization with a fixed source (i.e. boundary condtion) $J(t,\vec{x})$ while turning on an on-shell fluctuation mode $\delta\phi(z,\vec{x})$ which satisfies
\begin{equation}
\delta\phi(x)\sim z^{\Delta_{+}}J'(t,\vec{x})\,,\quad\text{as $z\rightarrow0$}\,,\label{eq:osflbg}
\end{equation}
with the boundary condition Equ.~(\ref{eq:fbcs}) automatically satisfied. Then we evaluate the resulting on-shell action (or perform the path integral with the prescribed boundary condition Equ.~(\ref{eq:bcs})) we get
\begin{equation}
\begin{split}
  Z[J,J']_{\text{CFT}}&=e^{-S^{\text{on-shell}}[\phi_{J}+\delta\phi]}\\&=e^{i\pi^{\frac{d}{2}}\frac{\Gamma[\Delta_{+}+1-\frac{d}{2}]}{\Gamma[\Delta_{+}]}\int dtd^{d-1}\vec{x}dt'd^{d-1}\vec{x'}\frac{J(t',\vec{x'})J(t,\vec{x})}{(-(t-t')^2+|\vec{x}-\vec{x'}|^{2})^{\Delta_{+}}}+i\pi^{\frac{d}{2}}\frac{\Gamma[\Delta_{+}+1-\frac{d}{2}]}{\Gamma[\Delta_{+}]}\int dtd^{d-1}\vec{x} J(t,\vec{x})J'(t,\vec{x})}\,,
  \end{split}
\end{equation}
where we ignored the quantum correction which is not relevant to the CFT generating functional. Finally, we integrate over the boundary source $J(\vec{x})$ and get Equ.~(\ref{eq:altgen}).

In summary, at the level of the partition function, the alternative quantization can be obtained from the standard quantization by turning on a profile of an on-shell mode Equ.~(\ref{eq:osflbg}) as a new piece of the background on top of $\phi_{J}$ and additionally doing the path integral also over the boundary source $J(\vec{x})$ (see \cite{Fichet:2021xfn} for a similar decomposition of the bulk field). This directly raises the question that what the fall-off behaviors of the on-shell and off-shell modes are in the alternative quantization scheme. 


In the rest of this section, we will see that the formalism we used for the discussion of the standard quantization can be easily modified and applied for the analysis in the alternative quantization.


\subsubsection{Partition Function}
In this subsection we will show that the formalism we used in Sec.~\ref{sec:partitons} also works in the computation of the partition function (i.e. the generating functional for CFT correlators) for the alternative quantization precisely when $\Delta_{-}>\frac{d-2}{2}$.

Similar as before, we have to solve the equation of motion with a boundary condition $J'(\vec{x})$
\begin{equation}
  (\Box-m^2)\phi_{J'}(\vec{x})=0\,,
\end{equation}
and the boundary condition is specified by setting the variation of the whole action $\delta S_{0}+\delta S_{bdy}$ to zero. However, now in the alternative quantization scheme we will use the following boundary term
\begin{equation}
  S_{bdy}=-\epsilon^{-\Delta_{-}}\pi^{\frac{d}{2}}\frac{\Gamma[\Delta_{-}+1-\frac{d}{2}]}{\Gamma[\Delta_{-}]}\int dtd^{d-1}\vec{x}\phi(\epsilon,t,\vec{x})J'(t,\vec{x})\,,
\end{equation}
which gives the boundary condition $J'(t,\vec{x})$ from $\delta S_{0}+\delta S_{bdy}=0$:
\begin{equation}
  z^{-\Delta_{+}}(z\partial_{z}-\Delta_{-})\phi(x)|_{z=\epsilon\rightarrow0}=-2\pi^{\frac{d}{2}}\frac{\Gamma[\Delta_{-}+1-\frac{d}{2}]}{\Gamma[\Delta_{-}]}J'(t,\vec{x})\,,\label{eq:bca}
\end{equation}
for a fixed $J'(t,\vec{x})$ and a generic fluctuation should satisfy 
\begin{equation}
 z^{-\Delta_{+}} (z\partial_{z}-\Delta_{-})\delta\phi(x)=0\,,\quad\text{as $z\rightarrow0$}\,,\label{eq:fbca}
\end{equation}
where again we have defined $\Delta_{\pm}=\frac{d}{2}\pm\sqrt{\frac{d^2}{4}+m^2}$. 

A solution of the equation of motion $\phi_{J'}(x)$ which is regular in interior of the AdS$_{d+1}$ \cite{Witten:1998qj} and satisfies the boundary condition Equ.~(\ref{eq:bca}) is
\begin{equation}
\phi_{J'}(x)=\int dt'd^{d-1}\vec{x'}\frac{J'(t',\vec{x'})z^{\Delta_{-}}}{(z^2-(t-t')^2+|\vec{x}-\vec{x'}|^2)^{\Delta_{-}}}\,.\label{eq:solutiona}
\end{equation}
To check that Equ.~(\ref{eq:solutiona}) satisfies the boundary condition Equ.~(\ref{eq:bcs}) it is important to know that
\begin{equation}
 \lim_{z\rightarrow0}\frac{z^{2\Delta-d}}{(z^2-(t-t')^2+|\vec{x}-\vec{x'}|^2)^{\Delta}}=\pi^{\frac{d}{2}}\frac{\Gamma[\Delta-\frac{d}{2}]}{\Gamma[\Delta]}\delta(t-t')\delta^{(d-1)}(\vec{x}-\vec{x'})\,,\quad\text{for }\quad\Delta>\frac{d}{2}\,,
\end{equation}
which imposes the constraint\footnote{The expansion of Equ.~(\ref{eq:solutiona}) for small $z$ behaves as
\begin{equation}
\phi_{J'}(z,t,\vec{x})=\Big(A_{J'}(t,\vec{x})z^{\Delta-}+\mathcal{O}(z^{\Delta_{-}+2})\Big)+\Big(B_{J'}(t,\vec{x})z^{\Delta_{+}}+z^{\Delta_{+}+2}\Big)\,,
\end{equation}
where $A_{J'}$ and $B_{J'}$ are solely determined by $J'(t,\vec{x})$. To make sure that the boundary condition Equ.~(\ref{eq:bca}) is satisfied we have to make sure that the $\mathcal{O}(z^{\Delta_{-}+2})$ terms in the first bracket give zero contribution to the left hand side of Equ.~(\ref{eq:bca}). Hence we need
\begin{equation}
 z^{\Delta_{-}+2-\Delta_{+}}=z^{-d+2+2\Delta_{-}}\rightarrow0\,,\quad\text{as}\quad z\rightarrow0\,.
\end{equation}
This requires that $\Delta_{-}>\frac{d-2}{2}$ which is exactly the same as the constraint Equ.~(\ref{eq:cwa}).}
\begin{equation}
 \Delta_{-}>\frac{d-2}{2}\,.\label{eq:cwa}
\end{equation}

Now we can evaluate the on-shell action
\begin{equation}
  S^{\text{on-shell}}[\phi_{J}]=S_{0}[\phi_{J}]+S_{bdy}[\phi_{J}]=-\pi^{\frac{d}{2}}\frac{\Gamma[\Delta_{-}+1-\frac{d}{2}]}{\Gamma[\Delta_{-}]}\int d^{d}\vec{x}d^{d}\vec{x'}\frac{J'(\vec{x'})J'(\vec{x})}{|\vec{x}-\vec{x'}|^{2\Delta_{-}}}\,,
\end{equation}
and then we get the expected CFT$_{d}$ generating functional
\begin{equation}
  Z[J']_{\text{CFT}}=e^{-i\pi^{\frac{d}{2}}\frac{\Gamma[\Delta_{-}+1-\frac{d}{2}]}{\Gamma[\Delta_{-}]}\int dtd^{d-1}\vec{x}dt'd^{d-1}\vec{x'}\frac{J'(t',\vec{x'})J'(t,\vec{x})}{(-(t-t')^2+|\vec{x}-\vec{x'}|^{2})^{\Delta_{-}}}}\,,\label{eq:treeZa}
\end{equation}
which is that for correlators of a conformal primary operator of conformal weight $\Delta_{-}$.

\subsubsection{On-Shell and Off-Shell Modes}\label{sec:altpr}
In analogy to Sec.~\ref{sec:prop}, we will study the properties of the on-shell and off-shell modes near the asymptotic boundary $z\rightarrow0$.

The on-shell modes satisfy the equation of motion
\begin{equation}
 (\Box-m^2)\delta\phi(x)=0\,,\label{eq:eomfa}
\end{equation}
for which a general solution is given by
\begin{equation}
 \delta\phi(z,\omega,\vec{k})^{\pm}=e^{-i\omega t+i \vec{k}\cdot\vec{x}}z^{\frac{d}{2}}J_{\pm\sqrt{\frac{d^2}{4}+m^2}}(z\sqrt{\omega^2-k^2})\,,\label{eq:afl}
\end{equation}
 where $\omega^{2}-\vec{k}^2\geq0$ and the fluctuation $\delta\phi^{\pm}$ has the asymptotic behavior
\begin{equation}
  \delta\phi^{\pm}(z,\omega,\vec{k})\sim z^{\Delta_{\pm}},\quad\text{as}\quad z\rightarrow0\,.
\end{equation}
Moreover the Klein-Gordon norm 
\begin{equation} 
\begin{split}(\delta\phi^{\pm},\delta\phi^{\pm})_{\text{KG}}&=i\int_{\Sigma} dz d^{d-1}x\sqrt{-g}g^{tt}(\delta\phi^{\pm*}\partial_{t}\delta\phi^{\pm}-\delta\phi^{\pm}\partial_{t}\delta\phi^{\pm*})(z,\vec{x})\\
&\sim \int dz z^{-d+1}z^{2\Delta_{\pm}}\,,\quad\text{near}\quad z=0\,,
\end{split}
\end{equation}
is finite for both $\Delta_{\pm}$ if $\Delta_{-}>\frac{d-2}{2}$. As opposed to the standard quantization, the alternative quantization projects out the $\Delta_{+}$ modes. Hence the on-shell modes satisfy
\begin{equation}
\delta\phi_{\text{on-shell}}(x)\sim z^{\Delta_{-}}\,,\quad\text{as $z\rightarrow0$}\,.
\end{equation}

Let's now study the off-shell modes. Similar to the standard quantization case, Equ.~(\ref{eq:fbca}) should be satisfied for the stability of the on-shell configuration. This projects out the $\delta\phi^{-}$ modes and requires that $\lambda>0$. Moreover the normalizability of the measure Equ.~(\ref{eq:offshell2}) and the vanishing of 
\begin{equation}
S_{bdy}[\delta\phi]=-\epsilon^{-\Delta_{+}}\pi^{\frac{d}{2}}\frac{\Gamma[\Delta_{+}+1-\frac{d}{2}]}{\Gamma[\Delta_{+}]}\int d^{d}\vec{x}\delta\phi(\epsilon,\vec{x})J'(\vec{x})\,,
\end{equation}
as $\epsilon\rightarrow0$ are trivially satisfied for $\delta\phi^{+}$ modes. Hence in the alternative quantization the off-shell modes are constrained to satisfy
\begin{equation}
\delta\phi_{\text{off-shell}}(x)\rightarrow z^{\Delta_{+}(\lambda)}\,,\quad\text{as $z\rightarrow0$}\,,\label{eq:offshellal}
\end{equation}
and $\lambda>0$.
 
With the knowledge of the on-shell modes, we can construct the Hilbert space by doing the canonical quantization Equ.~(\ref{eq:canonicalqu}). The field operator $\hat{\phi}$(x) is given by
\begin{equation}
\hat{\phi}(x)=\int\frac{d\vec{k}}{(2\pi)^{d-1}}\int_{\omega^2-\vec{k}^2\geq0}\frac{d\omega}{2\pi}\Big(\delta\phi(x)_{\omega,\vec{k}}^{-}\hat{b}^{\dagger}_{\omega,\vec{k}}+\delta\phi(x)^{-*}_{\omega,\vec{k}}\hat{b}_{\omega,\vec{k}}\Big)\,,\label{eq:canonicala}
\end{equation}
where $\hat{b}^{\dagger}_{\omega,\vec{k}}$ and $\hat{b}_{\omega,\vec{k}}$ are the creation and annihilation operators. They satisfy the standard commutation relations
\begin{equation}
[\hat{b}_{\omega,\vec{k}},\hat{b}^{\dagger}_{\omega,\vec{k'}}]=\delta(\omega-\omega')\delta^{d-1}(\vec{k}-\vec{k'})\,.
\end{equation}

With the knowledge of the off-shell modes, we can understand the quantum correction to the partition function Equ.~(\ref{eq:treeZa}). A general field configuration can be expanded as
\begin{equation}
\begin{split}
\phi(x)=&\phi_{J'}(x)+\sum_{\lambda>0}\int_{\omega^2-\vec{k}^2\geq0}  \frac{d\vec{k}}{(2\pi)^{d-1}}\int d\omega b_{\lambda,\omega,\vec{k}}\delta\phi_{\lambda}(z,\omega,\vec{k})e^{-i\omega t+i\vec{k}\cdot \vec{x}}\,,\label{eq:almode}
\end{split}
\end{equation}
and the action evaluated on this configuration is
\begin{equation}
\begin{split}
S[\phi]&=S_{0}[\phi]+S_{bdy}[\phi]\\
&=-\pi^{\frac{d}{2}}\frac{\Gamma[\Delta_{-}+1-\frac{d}{2}]}{\Gamma[\Delta_{-}]}\int dtd^{d-1}\vec{x}dt'd^{d-1}\vec{x'}\frac{J'(t',\vec{x'})J'(t,\vec{x})}{(-(t-t')^2+|\vec{x}-\vec{x'}|^{2})^{\Delta_{-}}}\\
&\qquad +\frac{1}{2}\sum_{\lambda>0}\int_{\omega^2-\vec{k}^2\geq0} \frac{d^{d-1}\vec{k}}{(2\pi)}d\omega\lambda|b_{\lambda, \omega,\vec{k}}|^2\,.
\end{split}
\end{equation}
The first term gives us the result Equ.~(\ref{eq:treeZa}) and the second term describes the quantum fluctuations. The quantum correction to the partition function is obtained by the Gaussian integral over the coefficients $b_{\lambda,\omega,\vec{k}}$ as in Equ.~(\ref{eq:quantumZ}) and Equ.~(\ref{eq:1loop}).

\section{Consistency with the Double-trace Deformation}\label{sec:dbt}
We have performed a careful analysis of the bulk modes in the standard AdS/CFT correspondence in the previous section. In this section, we will review the description of double-trace deformation in the standard AdS/CFT correspondence and check the consistency with our previous analysis. For the sake of convenience, we will use the Euclidean signature from now on. Since we will always think of analytic continuation to the Lorentzian signature, we will ignore the subtleties of the Euclidean AdS/CFT such as the bulk irregularity of the $\Delta_{-}$ modes.

\subsection{Double-trace Deformation in AdS/CFT}
As it is proposed by Witten \cite{Witten:2001ua} that the double-trace deformation $:\int d^{d}\vec{x} \mathcal{O}^{2}(\vec{x}):$ (where for the sake of simplicity we will ignore the normal ordering hereafter) on the CFT side duals to the modification of the boundary condition for the scalar field $\phi(z,\vec{x})$ (which is dual to the CFT single-trace operator $\mathcal{O}(\vec{x})$) in the AdS bulk. More precisely, if the deformation is given by
\begin{equation}
 W[\mathcal{O}]=\frac{h}{2}\int d^{d}x \mathcal{O}^2(\vec{x})\,,\label{eq:dtdef}
\end{equation}
where the single-trace operator $\mathcal{O}(\vec{x})$ has conformal weight $\Delta=\Delta_{\pm}$ depending on our quantization scheme and $h$ is the coupling constant. Then the dual scalar field in AdS$_{d+1}$ has the near boundary behavior
\begin{equation}
\begin{split}
   \phi(z,\vec{x})&\sim \Big(\gamma_{\Delta}\frac{\delta W[\alpha]}{\delta\alpha} z^{d-\Delta}+\mathcal{O}(z^{d-\Delta+2})\Big)+\Big(\alpha(\vec{x})z^{\Delta}+\mathcal{O}(z^{\Delta+2})\Big)\,\\&= \Big(h\gamma_{\Delta}\alpha(\vec{x}) z^{d-\Delta}+\mathcal{O}(z^{d-\Delta+2})\Big)+\Big(\alpha(\vec{x})z^{\Delta}+\mathcal{O}(z^{\Delta+2})\Big),\label{eq:witten}
   \end{split}
\end{equation}
for a generic configuration with the boundary condition $J(t,\vec{x})=0$ where $\gamma_{\Delta}$ is an order one constant. We want to get the precise value of $\gamma_{\Delta}$ so we need a precise description of the double-trace deformation Equ.~(\ref{eq:dtdef}) in the dual AdS bulk.

The precise description can be obtained using the path integral formulation. The double-trace deformation can be formulated using the path integral language \cite{Berkooz:2002ug}. Let's consider the CFT generating functional under the double-trace deformation Equ.~(\ref{eq:dtdef}):
\begin{equation}
\begin{split}
  Z[J]_{\text{CFT}}&=\langle e^{\int d^{d}\vec{x}\Big(\frac{h}{2}\mathcal{O}^{2}(\vec{x})+J(\vec{x})\mathcal{O}(\vec{x})\Big)}\rangle_{\text{CFT}}=\langle \int D[\lambda(\vec{x})]e^{\int d^{d}\vec{x}\Big(-\frac{\lambda^{2}(\vec{x})}{2h}+(J(\vec{x})+\lambda(\vec{x}))\mathcal{O}(\vec{x})\Big)}\rangle_{\text{CFT}}\,\\&=\int D[\lambda(\vec{x})]e^{-\int d^{d}\vec{x}\frac{(\lambda-J)^{2}(\vec{x})}{2h}}\langle e^{\int d^{d}\vec{x}\lambda(\vec{x})\mathcal{O}(\vec{x})}\rangle_{\text{CFT}}\,\\&=\int D[\lambda(\vec{x})]e^{-\int d^{d}\vec{x}\frac{(\lambda-J)^{2}(\vec{x})}{2h}}Z[\lambda]_{AdS_{d+1}}=\int D[\lambda(\vec{x})]e^{-\int d^{d}\vec{x}\frac{(\lambda-J)^{2}(\vec{x})}{2h}}\int D[\phi;\lambda]e^{-S_{\lambda}[\phi]}\,\\&=\int D[\lambda(\vec{x})]e^{-\int d^{d}\vec{x}\frac{(\lambda-J)^{2}(\vec{x})}{2h}}\int D[\phi]e^{-S_{\lambda}[\phi]}\Pi_{\vec{x}}\delta\Big(\epsilon^{-d+\Delta}(z\partial_{z}-\Delta)\phi(x)+2\pi^{\frac{d}{2}}\frac{\Gamma[\Delta+1-\frac{d}{2}]}{\Gamma[\Delta]}\lambda(\vec{x})\Big)\,\\&=\int D[\lambda(\vec{x})]e^{-\int d^{d}\vec{x}\frac{(\lambda-J)^{2}(\vec{x})}{2h}}\int D[\phi]D[\beta(\vec{x})]e^{-S_{\lambda}[\phi]}e^{i\int d^{d}\vec{x}\beta(\vec{x})\Big(\epsilon^{-d+\Delta}(z\partial_{z}-\Delta)\phi(x)+2\pi^{\frac{d}{2}}\frac{\Gamma[\Delta+1-\frac{d}{2}]}{\Gamma[\Delta]}\lambda(\vec{x})\Big)}\,,
  \end{split}
\end{equation}
where the path integral measure $D[\phi;\lambda]$ is constrained by the boundary condition Equ.~(\ref{eq:bcs}) now with a source $\lambda(\vec{x})$, the path integral measure $D[\phi]$ is not constrained by any boundary condition and $S_{\lambda}[\phi]$ is given by
\begin{equation}
 S_{\lambda}[\phi]=S_{0}[\phi]-\epsilon^{-\Delta}\pi^{\frac{d}{2}}\frac{\Gamma[\Delta+1-\frac{d}{2}]}{\Gamma[\Delta]}\int d^{d}\vec{x}\phi(\epsilon,\vec{x})\lambda(\vec{x})\,.
\end{equation}
We can integrate out $\lambda(\vec{x})$ and get
\begin{equation}
  Z[J]_{\text{CFT}}=\int D[\phi]D[\beta(\vec{x})]e^{-S[\phi,\beta]}\,,
\end{equation}
and the resulting exact effective action is given by
\begin{equation}
\begin{split}
 S[\phi,\beta]&=S_{0}[\phi]-\int d^{d}\vec{x}\epsilon^{-\Delta}\pi^{\frac{d}{2}}\frac{\Gamma[\Delta+1-\frac{d}{2}]}{\Gamma[\Delta]}\phi(\vec{x}) J(\vec{x})+\frac{h}{2}\int d^{d}\vec{x}\Big(\pi^{\frac{d}{2}}\frac{\Gamma[\Delta+1-\frac{d}{2}]}{\Gamma[\Delta]}\Big)^2\Big(4\beta(\vec{x})^{2}-\epsilon^{-2\Delta}\phi(\vec{x})^2\Big)\,\\&-i\int d^{d}x\beta(\vec{x})\Big(\epsilon^{-d+\Delta}(z\partial_{z}-\Delta)\phi(x)+2\pi^{\frac{d}{2}}\frac{\Gamma[\Delta+1-\frac{d}{2}]}{\Gamma[\Delta]}J(\vec{x})+2h(\pi^{\frac{d}{2}}\frac{\Gamma[\Delta+1-\frac{d}{2}]}{\Gamma[\Delta]})^{2}\epsilon^{-\Delta}\phi(\vec{x})\Big)\,.
 \end{split}
\end{equation}
Moreover, we can integrate out $\beta(\vec{x})$ and finally get
\begin{equation}
 Z[J]_{\text{CFT}}=\int D[\phi]e^{-S_{h}[\phi]}\,,
\end{equation}
where the exact effective action is given by
\begin{equation}
\begin{split}
S_{h}[\phi]&=S_{0}[\phi]-\int d^{d}\vec{x}\epsilon^{-\Delta}\pi^{\frac{d}{2}}\frac{\Gamma[\Delta+1-\frac{d}{2}]}{\Gamma[\Delta]}\phi(\vec{x}) J(\vec{x})-\frac{h}{2}\int d^{d}\vec{x}\Big(\pi^{\frac{d}{2}}\frac{\Gamma[\Delta+1-\frac{d}{2}]}{\Gamma[\Delta]}\Big)^{2}\epsilon^{-2\Delta}\phi(\vec{x})^2\\&+\frac{1}{2h}\int d^{d}\vec{x}\Big(2\pi^{\frac{d}{2}}\frac{\Gamma[\Delta+1-\frac{d}{2}]}{\Gamma[\Delta]}\Big)^{-2}\Big(\epsilon^{-d+\Delta}(z\partial_{z}-\Delta)\phi(x)+2\pi^{\frac{d}{2}}\frac{\Gamma[\Delta+1-\frac{d}{2}]}{\Gamma[\Delta]}J(\vec{x})\\&+2h(\pi^{\frac{d}{2}}\frac{\Gamma[\Delta+1-\frac{d}{2}]}{\Gamma[\Delta]})^{2}\epsilon^{-\Delta}\phi(\vec{x})\Big)^2\,,\label{eq:exacteffact}
\end{split}
\end{equation}
which reproduces our previous results
\begin{equation}
\begin{split}
S=S_{0}[\phi]-\int d^{d}\vec{x}\epsilon^{-\Delta}\pi^{\frac{d}{2}}\frac{\Gamma[\Delta+1-\frac{d}{2}]}{\Gamma[\Delta]}\phi(\vec{x}) J(\vec{x})\,,\\
\epsilon^{-d+\Delta}(z\partial_{z}-\Delta)\phi(\epsilon,\vec{x})=-2\pi^{\frac{d}{2}}\frac{\Gamma[\Delta+1-\frac{d}{2}]}{\Gamma[\Delta]}J(\vec{x})\,,
\end{split}
\end{equation}
when $h\rightarrow0$.

Now we want to derive the boundary condition for a generic field configuration when $J(t,\vec{x})=0$. This is ensured by the following requirement
\begin{equation}
\delta S_{h}[\phi]=-\int \sqrt{g}d^{d+1}x\delta\phi(\Box-m^2)\phi\,,
\end{equation}
which equivalently says that the boundary contribution of the variation is zero. A generic configuration of $\phi(x)$ has both on-shell and off-shell piece and as is known from our previous study that the off-shell piece decays faster than the on-shell piece as $z\rightarrow0$ so we only have to ensure the on-shell piece satisfies the above condition. A generic on-shell profile satisfies
\begin{equation}
\phi(x)\sim(\alpha(\vec{x}) z^{d-\Delta}+\mathcal{O}(z^{\Delta-d+2}))+(\beta\vec{x})(z^{\Delta}+\mathcal{O}(z^{\Delta+2}))\,,\quad\text{as $z\rightarrow0$}\,,
\end{equation}
and as long as we are in the window that we are able to perform the alternative quantization i.e. $\frac{d+2}{2}>\Delta>\frac{d-2}{2}$ we don't have to care about the subleading term so we can just use
\begin{equation}
\phi(x)\sim \alpha(\vec{x})z^{d-\Delta}+\beta(\vec{x})z^{\Delta}\,,\quad\text{as $z\rightarrow0$}\,.
\end{equation}
Now we can compute the boundary contribution of the variation $\delta S_{h}[\phi]$ for $J(t,\vec{x})=0$:
\begin{equation}
\delta S_{h}[\phi]_{bdy}=\int d^{d}\vec{x}\Big[\frac{1}{h}(2\pi^{\frac{d}{2}}\frac{\Gamma[\Delta+1-\frac{d}{2}]}{\Gamma[\Delta]})^{-2}(d-2\Delta)^{2}\alpha\delta\alpha+(d-2\Delta)\alpha\delta\alpha \epsilon^{d-2\Delta}+(d-2\Delta)\beta\delta\alpha\Big]\,.
\end{equation}
Since either $\epsilon^{d-2\Delta}$ or $\epsilon^{2\Delta-d}$ is divergent (as $\Delta=\frac{d}{2}\pm\frac{1}{2}\sqrt{d^2+4m^2}$), we need holographic renormalization \cite{Skenderis:2002wp} to remove the divergent one. Let's discussion the two situations separately.
\begin{itemize}
    \item $\Delta>\frac{d}{2}$: In this case, the counterterm we should add is 
    \begin{equation}
S_{\text{counter}}[\phi]=-\frac{d-2\Delta}{2}\int d^{d}\vec{x}\frac{1}{\epsilon^{d}}\phi^{2}(\epsilon,\vec{x})\,.\label{eq:counter1}
    \end{equation}
    Hence we have
    \begin{equation}
\delta S_{h}[\phi]_{bdy}+\delta S_{\text{counter}}[\phi]=\int d^{d}\vec{x}\Big[\frac{1}{h}(2\pi^{\frac{d}{2}}\frac{\Gamma[\Delta+1-\frac{d}{2}]}{\Gamma[\Delta]})^{-2}(d-2\Delta)^2\alpha\delta\alpha-2(d-2\Delta)\alpha\delta\beta\Big]=0\,,
    \end{equation}
which implies that
    \begin{equation}
\alpha=\frac{8h}{d-2\Delta}\Big(\pi^{\frac{d}{2}}\frac{\Gamma[\Delta+1-\frac{d}{2}]}{\Gamma[\Delta]}\Big)^{2}\beta\,.\label{eq:ab1}
    \end{equation}
\item $\Delta<\frac{d}{2}$: In this case, we just have to add a zero counterterm.
 Hence we have
    \begin{equation}
\delta S_{h}[\phi]_{bdy}+\delta S_{\text{counter}}[\phi]=\int d^{d}\vec{x}\Big[\frac{1}{h}(2\pi^{\frac{d}{2}}\frac{\Gamma[\Delta+1-\frac{d}{2}]}{\Gamma[\Delta]})^{-2}(d-2\Delta)^2\alpha\delta\alpha+(d-2\Delta)\beta\delta\alpha\Big]=0\,,
    \end{equation}
    which implies that
    \begin{equation}
\alpha=-\frac{4h}{d-2\Delta}\Big(\pi^{\frac{d}{2}}\frac{\Gamma[\Delta+1-\frac{d}{2}]}{\Gamma[\Delta]}\Big)^{2}\beta\,.\label{eq:ab2}
    \end{equation}
\end{itemize}

Now let's understand what's going on with the Hilbert space when we have the double-trace deformation for which we go back to Lorentzian signature. Depending on the quantization scheme that we start with we either have the boundary behavior Equ.~(\ref{eq:ab1}) or Equ.~(\ref{eq:ab2}). Without loss of generality, let's consider Equ.~(\ref{eq:ab1}). In this case, a general bulk on-shell mode is given by
\begin{equation}
\delta\phi(x)_{\omega,\vec{k}}=e^{-i\omega t+i \vec{k}\cdot\vec{x}}z^{\frac{d}{2}}\Big(\frac{8h}{d-2\Delta}(\pi^{\frac{d}{2}}\frac{\Gamma[\Delta+1-\frac{d}{2}]}{\Gamma[\Delta]})^{2}J_{-\sqrt{\frac{d^2}{4}+m^2}}(z\sqrt{\omega^2-k^2})+J_{\sqrt{\frac{d^2}{4}+m^2}}(z\sqrt{\omega^2-k^2})\Big)\,,\label{eq:dtmode}
\end{equation}
and these modes are orthonormal in the Klein-Gordon norm due to the orthonormality of the $\delta\phi^{\pm}(x)_{\omega,\vec{x}}$ in Equ.~(\ref{eq:sfl}). The field operator $\hat{\phi}(x)$ is given by
\begin{equation}
\hat{\phi}(x)=\int_{\omega^2-\vec{k}^2\geq0} \frac{d\vec{k}}{(2\pi)^{d-1}}\frac{d\omega}{2\pi}\Big(\delta\phi(x)_{\omega,\vec{k}}a^{\dagger}_{\omega,\vec{k}}+\delta\phi(x)_{\omega,\vec{k}}a_{\omega,\vec{k}}\Big)\,,\label{eq:dtcanonical}
\end{equation}
where the creation and annihilation operators satisfy the standard commutation relations if we impose the canonical quantization condition Equ.~(\ref{eq:canonicalqu}). Hence we see that the Hilbert space dimension is not changed but the asymptotic behavior of each eigenmode is deformed by the double-trace deformation Equ.~(\ref{eq:dtdef}). 
\subsection{Interpolation between the Two Quantization Schemes through the Double-trace Deformation}
We can see from Equ.~(\ref{eq:witten}) that in the strong coupling limit $h\rightarrow\infty$ we go from the quantization corresponding to $\Delta$ to that corresponding to $d-\Delta$. In other words, the double-trace deformation Equ.~(\ref{eq:dtdef}) drives the standard quantization to the alternative quantization and vice versa. This can be seen more precisely from the path integral point of view.
Now we can take the strong coupling limit $h\rightarrow\infty$ for Equ.(\ref{eq:exacteffact}). To have a legitimate large $h$ limit the relevant term has to be of order $\mathcal{O}(1)$ when we take $\epsilon\rightarrow0$ before we take the large $h$ limit. In other words we have
\begin{equation}
 \epsilon^{-d+\Delta}(z\partial_{z}-\Delta)\phi(x)+2\pi^{\frac{d}{2}}\frac{\Gamma[\Delta+1-\frac{d}{2}]}{\Gamma[\Delta]}J(\vec{x})\lesssim 1\,,\quad\text{as}\quad z\rightarrow0\,,
\end{equation}
which means that \footnote{Here we have used the CFT unitarity bound that $\Delta>\frac{d-2}{2}$ such that the $z^{\Delta}(1+\mathcal{O}(z^2))$ series contributes zero as $z=\epsilon\rightarrow0$.}
\begin{equation}
\phi(z,\vec{x})\lesssim z^{d-\Delta}\,.\label{eq:dtasym}
\end{equation}
Then the $h$-dependent terms in $S_{h}[\phi]$ goes to zero as $h\rightarrow\infty$. Hence we finally get
\begin{equation}
\int D[\phi]e^{-S_{0}[\phi]+\frac{1}{2}\int d^{d}\vec{x}\epsilon^{-d}\phi(\epsilon,\vec{x})(z\partial_{z}-\Delta)\phi(\epsilon,\vec{x})-S_{\text{counter}}[\phi]}\,.\label{eq:resultpi}
\end{equation}

Now we can see that if we start with the alternative quantization i.e. $\Delta=\Delta_{-}$ then the on-shell modes of the resulting path integral Equ.~(\ref{eq:resultpi}) in the Lorentzian signature satisfy $\delta\phi(z,\vec{x})\sim z^{d-\Delta_{-}}=z^{\Delta_{+}}$. The off-shell modes are those fluctuations (on top of an on-shell configuration) which satisfy Equ.~(\ref{eq:dtasym}) and has a finite action (here $\phi=\phi_{os}+\delta\phi$ where $\phi_{os}$ is an on-shell configuration)
\begin{equation}
 S[\phi]=S_{0}[\phi]-\frac{1}{2}\int d^{d}\vec{x}\epsilon^{-d}\phi(\epsilon,\vec{x})(z\partial_{z}-\Delta_{-})\phi(\epsilon,\vec{x})
\end{equation}
This is automatically satisfied if $\delta\phi(z,\vec{x})$ decays faster than $z^{\Delta_{+}}$ which is consistent with Equ.~(\ref{eq:offshellst}). Hence, we can see that if we start with the alternative quantization then we end up with the bulk modes as in the case of the standard quantization Sec.~\ref{sec:prop} if we take $h\rightarrow\infty$.

On the other hand, if we start with the standard quantization i.e. $\Delta=\Delta_{+}$ then from Equ.~(\ref{eq:dtasym}) the modes at least satisfy
\begin{equation}
\delta \phi(z,\vec{x})\sim z^{d-\Delta_{+}}=z^{\Delta_{-}}\,.
\end{equation}
Hence the on-shell modes satifies $\delta\phi(z,\vec{x})\sim z^{\Delta_{-}}$. For the off-shell modes (on top of an on-shell configuration) the finiteness of the action (here $\phi=\phi_{os}+\delta\phi$ where $\phi_{os}$ is an on-shell configuration)
\begin{equation}
S[\phi]=S_{0}[\phi]+\frac{1}{2}\int d^{d}\vec{x}\epsilon^{-d}\phi(\epsilon,\vec{x})(z\partial_{z}-\Delta_{+})\phi(\epsilon,\vec{x})-\frac{d-\Delta_{+}}{2}\int d^{d}\vec{x}\frac{1}{\epsilon^{d}}\phi^{2}(\epsilon,\vec{x})\,,\label{eq:div}
\end{equation}
requires that $\delta\phi(z,\vec{x})$ to decay faster than or equal to $z^{\Delta_{+}}$ which is consistent with Equ.~(\ref{eq:offshellal}). Therefore, we conclude that if we start with the standard quantization then we end up with the bulk modes as in the case of the alternative quantization Sec.~\ref{sec:altpr} if we take $h\rightarrow\infty$.

In summary, we see that our result of the bulk modes in Sec.~\ref{sec:prop} and Sec.~\ref{sec:altpr} are consistent with the fact that the double-trace deformation is interpolating between the standard quantization and alternative quantization in AdS/CFT.

\section{Coupling AdS/CFT to a Bath}\label{sec:dbto}
Having reviewed and carefully analyzed the physics of the standard AdS/CFT correspondence, we will study the main subject of this paper that we couple the AdS/CFT to a bath at the asymptotic boundary of the bulk AdS. The main techniques we will use are almost the same as before. We assume that we are in the window where we are able to perform the alternative quantization.

As we briefly mentioned in the introduction that the coupling protocol is most easily described in the CFT description. Let's consider the CFT$_{d}$ which duals to the gravitational theory in the bulk AdS$_{d+1}$ and let's call it CFT$^{1}$. We want to couple this CFT$_{d}$ to a bath which is modelled by another conformal field theory which we call CFT$^{2}$. We will consider two cases for the CFT$^{2}$.
\subsection{Non-Gravitational Bath}\label{sec:nongravbath}
In the first case, CFT$^{2}$ is of dimension $(d+1)$ and it lives on a half Minkowski space whose boundary is of the same geometry as the manifold that supports CFT$^{1}$ (see Fig.\ref{pic:nongravbath}). These two CFT's are coupled to each other as following
\begin{equation}
S_{\text{tot}}=S_{\text{CFT}^{1}_{d}}+g\int d^{d}\vec{x}:\mathcal{O}_{1}\mathcal{O}_{2}:+S_{\text{CFT}^{2}_{d+1}}\,,
\end{equation}
where $\mathcal{O}_{1}$ is a single-trace operator from CFT$^{1}_{d}$, $\mathcal{O}_{2}$ is the boundary extrapolation of a single-trace operator of CFT$^{2}_{d+1}$ and the deformation is marginal. Hence we have $\Delta_{1}+\Delta_{2}=d$. We consider the case that $\Delta_{1}>\frac{d}{2}$ (i.e. standard quantization in the bulk AdS$_{d+1}$). From the dual AdS perspective, this is equivalent to gluing the asymptotic boundary of the AdS$_{d+1}$ to a $(d+1)$-dimensional bath (i.e., the CFT$^{2}$) and the gluing condition is determined by the double-trace deformation. Following \cite{Witten:2001ua} now we have
\begin{equation}
W[\mathcal{O}_{1},\mathcal{O}_{2}]=g\int d^{d}\vec{x}\mathcal{O}_{1}\mathcal{O}_{2}\,,
\end{equation}
and the scalar field in AdS$_{d+1}$ that duals to $\mathcal{O}_{1}$ has the near boundary behavior
\begin{equation}
\begin{split}
\phi(z,\vec{x})&\sim (a_{\Delta}\frac{\delta W[\alpha,\mathcal{O}_{2}]}{\delta\alpha}z^{d-\Delta_{1}}+\mathcal{O}(z^{d-\Delta_{1}+2}))+(\alpha(\vec{x})z^{\Delta_{1}}+\mathcal{O}(z^{\Delta_{1}+2}))\\&\sim(ga_{\Delta}\mathcal{O}_{2}(\vec{x})z^{d-\Delta_{1}}+\mathcal{O}(z^{d-\Delta_{1}+2}))+(\alpha(\vec{x})z^{\Delta_{1}}+\mathcal{O}(z^{\Delta_{1}+2}))\,,\label{eq:bcnong}
\end{split}
\end{equation}
where $\mathcal{O}_{2}(\vec{x})$ and $\alpha(\vec{x})$ are independent. To derive the proper action for $\phi(x)$ let's consider the source free partition function
\begin{equation}
\begin{split}
Z_{\text{CFT}^{1}+\text{CFT}^{2}}&=\langle e^{-\int d^{d}\vec{x} \mathcal{O}_{1}(\vec{x})\mathcal{O}_{2}(\vec{x})}\rangle_{\text{CFT}^{1}+\text{CFT}^{2}}\\&=\langle Z[\mathcal{O}_{2}]_{AdS_{d+1}}\rangle_{\text{CFT}^{2}}\\&=\langle \int D[\phi;\mathcal{O}_2]e^{-S_{\mathcal{O}_{2}}[\phi]}\rangle_{\text{CFT}^{2}}\,.
\end{split}
\end{equation}
For simplicity we assume that the CFT$^{2}$ just gives a dynamics for $\mathcal{O}_{2}^{ext}$ whose boundary extrapolation is $\mathcal{O}_{2}$ and we write it as $S[\mathcal{O}_{2}^{ext}]$. Thus we have
\begin{equation}
\begin{split}
Z_{\text{CFT}^{1}+\text{CFT}^{2}}&=\int D[\mathcal{O}_{2}^{ext}]D[\phi;\mathcal{O}_2]e^{-S_{\mathcal{O}_{2}}[\phi]-S[\mathcal{O}_{2}^{ext}]}\,.\label{eq:partitontot1}
\end{split}
\end{equation}
Now let's determine the appropriate boundary conditions and boundary terms by considering the variation of $S_{0}[\phi]+S[\mathcal{O}_{2}^{ext}]$ around an on-shell configuration $(\phi_{os},\mathcal{O}_{2,os})$
\begin{equation}
\begin{split}
\delta S_{0}[\phi]+\delta S[\mathcal{O}_{2}^{ext}]&=-\frac{1}{2}\int_{z=\epsilon} d^{d}\vec{x}z^{-d+1}\Big(\delta\phi\partial_{z}\phi_{os}-\phi_{os}\partial_{z}\delta\phi\Big)+\delta S[\mathcal{O}_{2,os}^{ext}]\,.
\end{split}
\end{equation}
Since the off-shell modes are subleading, we only have to focus on the on-shell piece for $\delta\phi$. Hence we have
\begin{equation}
\begin{split}
\delta S_{0}[\phi]+\delta S[\mathcal{O}_{2}^{ext}]&=-\frac{1}{2}ga_{\Delta}(2\Delta_{1}-d)\int d^{d}\vec{x}(\delta\mathcal{O}_{2}\alpha-\delta\alpha\mathcal{O}_{2})+\delta S[\mathcal{O}_{2,os}^{ext}]\,.
\end{split}
\end{equation}
To proceed we need information about $S[\mathcal{O}_{2}^{ext}]$. It is a sector of the bath CFT$_{d+1}$ and without loss of generality we assume that we have
\begin{equation}
 S[\mathcal{O}_{2}^{ext}]=\frac{1}{2}\Big(\int d^{d+1}x\mathcal{O}_{2}^{ext}\partial^{\mu}\partial_{\mu}\mathcal{O}_{2}^{ext}-V[\mathcal{O}_{2}^{ext}]\Big)\,,
\end{equation}
where the potential is added to ensure $\mathcal{O}_{2}^{ext}$ has the conformal dimension $\Delta_{2}=d-\Delta_{1}$ when extrapolated to the boundary and we assume that the direction normal to the boundary is $\perp$.\footnote{One can think of $\mathcal{O}_{2}^{ext}$ as a generalized free field.} Hence we have
\begin{equation}
\begin{split}
\delta S_{0}[\phi]+\delta S[\mathcal{O}_{2}^{ext}]&=-\frac{1}{2}ga_{\Delta}(2\Delta_{1}-d)\int d^{d}\vec{x}(\delta\mathcal{O}_{2}\alpha-\delta\alpha\mathcal{O}_{2})+\delta S[\mathcal{O}_{2,os}^{ext}]\\&=-\frac{1}{2}ga_{\Delta}(2\Delta_{1}-d)\int d^{d}\vec{x}(\delta\mathcal{O}_{2}\alpha-\delta\alpha\mathcal{O}_{2})+\frac{1}{2}\int d^{d}\vec{x} (\mathcal{O}_{2,os}\partial_{\perp}\delta\mathcal{O}_{2}^{ext}-\delta\mathcal{O}_{2}\partial_{\perp}\mathcal{O}_{2,os}^{ext})\,,
\end{split}
\end{equation}
where we ignored the contribution from the potential which is not a boundary term and this boundary variation is vanishing if
\begin{equation}
ga_{\Delta}=\frac{1}{(2\Delta_{1}-d)}\,\quad\alpha=-\partial_{\perp}\mathcal{O}_{2,os}^{ext}\,.\label{eq:a}
\end{equation}
Hence we don't need additional boundary terms on the action. Moreover, the off-shell modes of $\mathcal{O}_{2}^{ext}$ are taken to satisfy either the Dirichlet or the Neumann boundary condition.


Now let's understand the Hilbert space in the AdS$_{d+1}$ by doing canonical quantization Equ.~(\ref{eq:canonicalqu}) with the boundary condition Equ.~(\ref{eq:bcnong}). From this question we switch back to the Lorentzian signature. Equ.~(\ref{eq:bcnong}) tells us that we should treat $\delta\phi^{+}(x)_{\omega,\vec{k}}$ and $\delta\phi^{-}(x)_{\omega,\vec{k}}$ from Equ.~(\ref{eq:sfl}) independently and both of them qualify as on-shell modes so we have the field operator $\hat{\phi}(x)$
\begin{equation}
\hat{\phi}(x)=\int_{\omega^2-\vec{k}^2\geq0} \frac{d\vec{k}}{(2\pi)^{d-1}}\frac{d\omega}{2\pi}\Big(\delta\phi^{+}(x)_{\omega,\vec{k}}\hat{a}^{\dagger}_{\omega,\vec{k}}+\delta\phi^{-}(x)_{\omega,\vec{k}}\hat{b}^{\dagger}_{\omega,\vec{k}}+\delta\phi^{+*}(x)_{\omega,\vec{k}}\hat{a}_{\omega,\vec{k}}+\delta\phi^{-*}(x)_{\omega,\vec{k}}\hat{b}_{\omega,\vec{k}}\Big)\,,
\end{equation}
in which we have two copies of creation and annihilation operators. This tells us that the Hilbert space now is twice as large are before which is not surprising as we have coupled the AdS$_{d+1}$ to a bath so the particles from the bath are free to enter the AdS$_{d+1}$ \cite{Breitenlohner:1982jf}.

With the structure of the Hilbert space clear, we want to further understand the partition function Equ~(\ref{eq:partitontot1}) with the quantum correction included. The difference between this and the previous cases is that the boundary source $\mathcal{O}_{2}$ is also dynamical so the on-shell part of the field configuration is not fixed and it will be fluctuating. A generic bulk field configuration in AdS$_{d+1}$ can be expanded as
\begin{equation}
\begin{split}
\phi(z,t,\vec{x})=&\int_{\omega^2-\vec{k}^2\geq0}  \frac{d\vec{k}}{(2\pi)^{d-1}}\int d\omega \frac{1}{2\Delta_{1}-d}\mathcal{O}_{2,os,\omega,\vec{k}}\delta\phi^{-}(x)_{\omega,\vec{k}}+\int_{\omega^2-\vec{k}^2\geq0}  \frac{d\vec{k}}{(2\pi)^{d-1}}\int d\omega a_{0,\omega,\vec{k}}\delta\phi^{+}(x)_{\omega,\vec{k}}\\+&\sum_{\lambda>0}\int_{\omega^2-\vec{k}^2\geq0}  \frac{d\vec{k}}{(2\pi)^{d-1}}\int d\omega a_{\lambda,\omega,\vec{k}}\delta\phi_{\lambda}(z,\omega,\vec{k})e^{-i\omega t-i\vec{k}\cdot \vec{x}}\,.\label{eq:nongravmode}
\end{split}
\end{equation}
We can evaluate the action on the configuration as
\begin{equation}
\begin{split}
S_{0}[\phi]=\frac{1}{2}\sum_{\lambda>0}\int_{\omega^2-\vec{k}^2\geq0} \frac{d^{d-1}\vec{k}}{(2\pi)}d\omega\lambda|a_{\lambda, \omega,\vec{k}}|^2\,.\label{eq:acnongrav}
\end{split}
\end{equation}
where we notice that the zero modes decouple. Then the path integral can be evaluated as the integral over the coefficients $a_{\lambda,\omega,\vec{k}}$. Nevertheless, we also have to do the path integral over $\mathcal{O}_{2}$ for which again the zero modes $\mathcal{O}_{2,os}^{ext}$ will decouple from the kinetic term.

\subsection{Gravitational Bath}\label{sec:gravbath}
In the second case, CFT$^{2}$ is of dimension $d$ and it has a holographic dual (see Fig.\ref{pic:gravbath}). The two CFT's are again coupled to each other by a double-trace deformation
\begin{equation}
S_{\text{tot}}=S_{\text{CFT}^{1}_{d}}+g\int d^{d}\vec{x}:\mathcal{O}_{1}\mathcal{O}_{2}:+S_{\text{CFT}^{2}_{d}}\,,
\end{equation}
where $\mathcal{O}_{1}$ is a single-trace operator from CFT$^{1}$ and $\mathcal{O}_{2}$ is a single-trace operator from CFT$^{2}$. Hence physically we are coupling two AdS$_{d+1}$'s by gluing them along their asymptotic boundaries such that energy can flow between them.\footnote{See \cite{bintanja2023tunneling} for a study of gravitational thermodynamics of this model.} We will stick with this two-AdS$_{d+1}$ description. Now we have two scalar fields with one on each AdS$_{d+1}$ and they are dual to $\mathcal{O}_{1}$ and $\mathcal{O}_{2}$ respectively. Their near boundary behaviors are
\begin{equation}
\begin{split}
\phi_{1}(z,\vec{x})&\sim (a_{\Delta_{1}}\frac{\delta W[\alpha,\beta]}{\delta\alpha}z^{d-\Delta_{1}}+\mathcal{O}(z^{d-\Delta_{1}+2}))+(\alpha(\vec{x})z^{\Delta_{1}}+\mathcal{O}(z^{\Delta_{1}+2}))\\&\sim(\frac{1}{2\Delta_{1}-d}\beta(\vec{x})z^{d-\Delta_{1}}+\mathcal{O}(z^{d-\Delta_{1}+2}))+(\alpha(\vec{x})z^{\Delta_{1}}+\mathcal{O}(z^{\Delta_{1}+2}))\,,\\\phi_{2}(z,\vec{x})&\sim (a_{\Delta_{2}}\frac{\delta W[\alpha,\beta]}{\delta\beta}z^{d-\Delta_{2}}+\mathcal{O}(z^{d-\Delta_{2}+2}))+(\beta(\vec{x})z^{\Delta_{2}}+\mathcal{O}(z^{\Delta_{2}+2}))\\&\sim(\frac{1}{2\Delta_{2}-d}\alpha(\vec{x})z^{d-\Delta_{2}}+\mathcal{O}(z^{d-\Delta_{2}+2}))+(\beta(\vec{x})z^{\Delta_{2}}+\mathcal{O}(z^{\Delta_{2}+2}))\,,\label{eq:bcgr}
\end{split}
\end{equation}
where we have used the first formula in Equ.~(\ref{eq:a}) and we remember that $\Delta_{1}+\Delta_{2}=d$. We may have to add appropriate boundary terms to the action ensuring that this is a legitimate boundary condition. Let's firstly study the variation without any boundary counterterm around an on-shell configuration
\begin{equation}
\begin{split}
\delta S_{0}[\phi_{1}]+\delta S_{0}[\phi_{2}]&=-\frac{1}{2}\int_{z=\epsilon} d^{d}\vec{x}z^{-d+1}\Big(\delta\phi_{1}\partial_{z}\phi_{1,os}-\phi_{1,os}\partial_{z}\delta\phi_{1}\Big)-\frac{1}{2}\int_{z=\epsilon} d^{d}\vec{x}z^{-d+1}\Big(\delta\phi_{2}\partial_{z}\phi_{2,os}-\phi_{2,os}\partial_{z}\delta\phi_{2}\Big)\\&=-\frac{1}{2}\int d^{d}\vec{x}(\delta\beta\alpha-\delta\alpha\beta)-\frac{1}{2}\int d^{d}\vec{x}(\delta\alpha\beta-\delta\beta\alpha)\\&=0\,.
\end{split}
\end{equation}
Hence we don't need any boundary term added.

Now we go back to the Lorentzian signature. The construction of the Hilbert space in each AdS$_{d+1}$ is almost the same as in the nongravitational bath case. The field operators $\hat{\phi}_{1}(x)$ and $\hat{\phi}_{2}(x)$ can be expanded in terms of two sets of creation and annihilation operators as
\begin{equation}
\begin{split}
\hat{\phi}_{1}(x)&=\int_{\omega^2-\vec{k}^2\geq0} \frac{d\vec{k}}{(2\pi)^{d-1}}\frac{d\omega}{2\pi}\Big(\delta\phi^{+}(x)_{\omega,\vec{k}}\hat{a}^{\dagger}_{\omega,\vec{k}}+\frac{\delta\phi^{-}(x)_{\omega,\vec{k}}}{2\Delta_{1}-d}\hat{b}^{\dagger}_{\omega,\vec{k}}+\delta\phi^{+*}(x)_{\omega,\vec{k}}\hat{a}_{\omega,\vec{k}}+\frac{\delta\phi^{-*}(x)_{\omega,\vec{k}}}{2\Delta_{1}-d}\hat{b}_{\omega,\vec{k}}\Big)\,,\\
\hat{\phi}_{2}(x)&=\int_{\omega^2-\vec{k}^2\geq0} \frac{d\vec{k}}{(2\pi)^{d-1}}\frac{d\omega}{2\pi}\Big(\frac{\delta\phi^{+}(x)_{\omega,\vec{k}}}{2\Delta_{2}-d}\hat{a}^{\dagger}_{\omega,\vec{k}}+\delta\phi^{-}(x)_{\omega,\vec{k}}\hat{b}^{\dagger}_{\omega,\vec{k}}+\frac{\delta\phi^{+*}(x)_{\omega,\vec{k}}}{2\Delta_{2}-d}\hat{a}_{\omega,\vec{k}}+\delta\phi^{-*}(x)_{\omega,\vec{k}}\hat{b}_{\omega,\vec{k}}\Big)\,.\label{eq:canonicalgrav}
\end{split}
\end{equation}
We notice that we have
\begin{equation}
\begin{split}
[\hat{\phi}_{1}(z,t,\vec{x}),g^{tt}\partial_{t}\hat{\phi}_{1}(z',t,\vec{x'})]&=\frac{1+(\frac{1}{2\Delta_{1}-d})^2}{\sqrt{-g}}\delta(z-z')\delta^{d-1}(\vec{x}-\vec{x'})\,,\quad[\hat{\phi}_{2}(z,t,\vec{x}),g^{tt}\partial_{t}\hat{\phi}_{1}(z',t,\vec{x'})]=0\,,\\ [\hat{\phi}_{2}(z,t,\vec{x}),g^{tt}\partial_{t}\hat{\phi}_{2}(z',t,\vec{x'})]&=\frac{1+(\frac{1}{2\Delta_{2}-d})^2}{\sqrt{-g}}\delta(z-z')\delta^{d-1}(\vec{x}-\vec{x'})\,,\quad [\hat{\phi}_{1}(z,t,\vec{x}),g^{tt}\partial_{t}\hat{\phi}_{2}(z',t,\vec{x'})]=0\,,
\end{split}
\end{equation}
where the second column is due to $2\Delta_{1}-d=d-2\Delta_{2}$ and they ensure the locality (or the cluster decomposition) between the two AdS$_{d+1}$ universes.\footnote{The locality between $\phi_{1}(z,t,\vec{x})$ and $\phi_{2}(z',t',\vec{x'})$ can be seen by computing their commutator. The result is \begin{equation}
\begin{split}
\langle[\hat{\phi}_{1}(z,t,\vec{x}),\hat{\phi}_{2}(z',t',\vec{x}')]\rangle&\propto
\Im\Big(G_{\Delta_{2}}(t-t'-i\epsilon,\vec{x}-\vec{x'},z-z')-G_{\Delta_{1}}(t-t'-i\epsilon,\vec{x}-\vec{x'},z-z')\Big)\,,
\end{split}
\end{equation}
where $G_{\Delta}(t,\vec{x},z)$ is the AdS$_{d+1}$ Green function for the massive scalar field whose dual CFT operator has conformal dimension $\Delta$. This has been shown to be local with respect to the geodesic distance $\cosh^{-1}\frac{(z+z')^2+(\vec{x}-\vec{x'}^2-(t-t')^2)}{zz'}$ in \cite{Aharony:2005sh} using the global coordinates of AdS.}

The partition function can be similarly studied as before. A generic bulk fields configuration can be expanded as
\begin{equation}
\begin{split}
\phi_{1}(z,t,\vec{x})=&\int_{\omega^2-\vec{k}^2\geq0}  \frac{d\vec{k}}{(2\pi)^{d-1}}\int d\omega \frac{1}{2\Delta_{1}-d}b_{0,\omega,\vec{k}}\delta\phi^{-}(x)_{\omega,\vec{k}}+\int_{\omega^2-\vec{k}^2\geq0}  \frac{d\vec{k}}{(2\pi)^{d-1}}\int d\omega a_{0,\omega,\vec{k}}\delta\phi^{+}(x)_{\omega,\vec{k}}\\+&\sum_{\lambda>0}\int_{\omega^2-\vec{k}^2\geq0}  \frac{d\vec{k}}{(2\pi)^{d-1}}\int d\omega a_{\lambda,\omega,\vec{k}}\delta\phi_{\lambda}(z,\omega,\vec{k})e^{-i\omega t-i\vec{k}\cdot \vec{x}}\,,\\\phi_{2}(z,t,\vec{x})=&\int_{\omega^2-\vec{k}^2\geq0}  \frac{d\vec{k}}{(2\pi)^{d-1}}\int d\omega b_{0,\omega,\vec{k}}\delta\phi^{-}(x)_{\omega,\vec{k}}+\int_{\omega^2-\vec{k}^2\geq0}  \frac{d\vec{k}}{(2\pi)^{d-1}}\int d\omega \frac{1}{2\Delta_{2}-d}a_{0,\omega,\vec{k}}\delta\phi^{+}(x)_{\omega,\vec{k}}\\+&\sum_{\lambda>0}\int_{\omega^2-\vec{k}^2\geq0}  \frac{d\vec{k}}{(2\pi)^{d-1}}\int d\omega b_{\lambda,\omega,\vec{k}}\delta\phi_{\lambda}(z,\omega,\vec{k})e^{-i\omega t-i\vec{k}\cdot \vec{x}}\,.\label{eq:gravmode}
\end{split}
\end{equation}
The action evaluated on this configuration is
\begin{equation}
\begin{split}
S_{0}[\phi_{1}]+S_{0}[\phi_{2}]=-\frac{1}{2}\sum_{\lambda>0}\int_{\omega^2-\vec{k}^2\geq0} \frac{d^{d-1}\vec{k}}{(2\pi)}d\omega\lambda|a_{\lambda, \omega,\vec{k}}|^2-\frac{1}{2}\sum_{\lambda>0}\int_{\omega^2-\vec{k}^2\geq0} \frac{d^{d-1}\vec{k}}{(2\pi)}d\omega\lambda|b_{\lambda, \omega,\vec{k}}|^2\,,\label{eq:acgrav}
\end{split}
\end{equation}
where the on-shell modes decoupled completely.

\section{Path Integral and the Diffeomorphism Invariance}\label{sec:PIDI}
In the previous section we got the action for generic field configurations Equ.~(\ref{eq:acnongrav}) and Equ.~(\ref{eq:acgrav}). The partition functions can be obtained by doing the path integral over these configurations. Nevertheless, a subtlety is that the path integral measure is defined upto an ambiguous constant and we have to nomralize the path integral measure first to fix this overall constant which is potentially infinite \cite{Ginsparg:1988ui,Peskin:1995ev}. Since we are considering a gravitational theory in the AdS$_{d+1}$ this measure had better to be diffeomorphism invariant and if not there will be diffeomorphism anomaly. Generally for a scalar field on a $(d+1)$-dimensional curved spacetime background the diffeomorphism invariant path integral measure is defined by
\begin{equation}
\int D[\phi] e^{-i\int d^{d+1}x\sqrt{-g}\phi^2(x)}=1\,.
\end{equation}
In this section, we will show that in the appearance of the bath the situation is rather subtle and interesting.

\subsection{Gravitational Bath}\label{sec:gravbathana}

\begin{figure}
    \centering
    \begin{tikzpicture}
       \draw[-,very thick,red](0,-2) to (0,2);
       \draw[fill=yellow, draw=none, fill opacity = 0.1] (0,-2) to (3,-2) to (3,2) to (0,2);
           \draw[-,very thick,red](0,-2) to (0,2);
       \draw[fill=blue, draw=none, fill opacity = 0.1] (0,-2) to (-3,-2) to (-3,2) to (0,2);
       \node at (-2,0)
       {\textcolor{black}{$AdS_{d+1}$}};
        \node at (2,0)
       {\textcolor{black}{$AdS_{d+1}$}};
       \draw [-{Computer Modern Rightarrow[scale=1.25]},thick,decorate,decoration=snake] (-1,-1) -- (1,-1);
       \draw [-{Computer Modern Rightarrow[scale=1.25]},thick,decorate,decoration=snake] (1,1) -- (-1,1);
       \draw[-,very thick,red] (9,-2) to (9,2);
       \draw[fill=red, draw=none, fill opacity = 0.1] (9,-2) to (5,-2) to (5,2) to (9,2);
       \draw [-{Computer Modern Rightarrow[scale=1.25]},thick,decorate,decoration=snake] (8,-1) -- (9,-1);
       \draw [-{Computer Modern Rightarrow[scale=1.25]},thick,decorate,decoration=snake] (9,1) -- (8,1);
        \node at (6,0)
       {\textcolor{black}{$AdS_{d+1}$}};
    \end{tikzpicture}
    \caption{\textbf{Left:}The situation we consider in Fig.\ref{pic:gravbath} where we have two AdS$_{d+1}$ universes coupled to each other by gluing them along their common asymptotic boundary using a double-trace deformation Equ.~(\ref{eq:gravdoubet}). We have one scalar field on each AdS$_{d+1}$ and they have transparent boundary condition at the asymptotic boundary due to the double-trace deformation. \textbf{Right:} For the sake of convenience we folded the two AdS$_{d+1}$ to a single AdS$_{d+1}$ where we have two scalar fields and the linearly independent combinations Equ.~(\ref{eq:irre}) have the reflective boundary condtion. Moreover, in the AdS$_{d+1}$ resulted from folding we have two dynamical gravitons.}
    \label{pic:gravbathfold}
\end{figure}
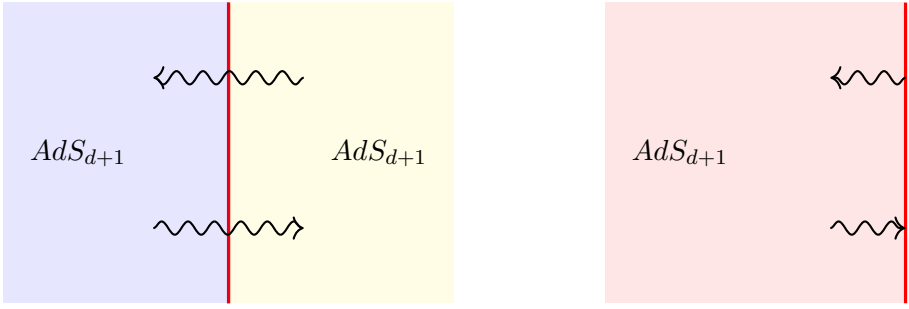

For the sake of simplicity, we consider first the case of the gravitational bath that we have studied in Sec.~\ref{sec:gravbath}. In Sec.~\ref{sec:gravbath} we noticed that the scalar fields $\phi_{1}(x)$ and $\phi_{2}(x)$ have rather unusual boundary conditions as comparing to the standard AdS/CFT case and a direct consequence of these unusual boundary conditions is that in the canonical quantization of them we need two sets of creation and annihilation operators for both of them.\footnote{Here we emphasize that they share the same two sets (see Equ.~(\ref{eq:canonicalgrav})).} This suggests that none of them is in an irreducible representation of the AdS$_{d+1}$ isometry group $SO(d,2)$. Nevertheless, specific linear combinations of them are in irreducible representations and they are
\begin{equation}
\begin{split}
(2\Delta_{1}-d)\hat{\phi}_{1}(x)-\hat{\phi}_{2}(x)&=(2\Delta_{1}-d-\frac{1}{2\Delta_{2}-d})\int_{\omega^2-\vec{k}^2\geq0} \frac{d\vec{k}}{(2\pi)^{d-1}}\frac{d\omega}{2\pi}\Big(\delta\phi^{+}(x)_{\omega,\vec{k}}\hat{a}^{\dagger}_{\omega,\vec{k}}+\delta\phi^{+*}(x)_{\omega,\vec{k}}\hat{a}_{\omega,\vec{k}}\Big)\,,\\
\hat{\phi}_{1}(x)-(2\Delta_{2}-d)\hat{\phi}_{2}(x)&=(\frac{1}{2\Delta_{1}-d}-2\Delta_{2}+d)\int_{\omega^2-\vec{k}^2\geq0} \frac{d\vec{k}}{(2\pi)^{d-1}}\frac{d\omega}{2\pi}\Big(\delta\phi^{-}(x)_{\omega,\vec{k}}\hat{b}^{\dagger}_{\omega,\vec{k}}+\delta\phi^{-*}(x)_{\omega,\vec{k}}\hat{b}_{\omega,\vec{k}}\Big)\,.\label{eq:irre}
\end{split}
\end{equation}
This fact can be more intuitively understood by folding the two AdS$_{d+1}$ to a single AdS$_{d+1}$ in which we would have two scalar fields $\phi_{1}(x)$ and $\phi_{2}(x)$ (and also two gravitons). Moreover, the stress-energy tensors of the two combinations in Equ.~(\ref{eq:irre}) have reflective boundary conditions on the asymptotic boundary of the resulting AdS$_{d+1}$(see Fig.\ref{pic:gravbathfold}). More precisely, in the resulting AdS$_{d+1}$, we are performing the standard quantization in AdS/CFT for $(2\Delta_{1}-d)\phi_{1}(x)-\phi_{2}(x)$ and we are performing the alternative quantization  for $\phi_{1}(x)-(2\Delta_{2}-d)\phi_{2}(x)$ (see Equ.~(\ref{eq:canonicals}) and Equ.~(\ref{eq:canonicala})). Hence the path integral of the scalar fields can be better studied in the basis $\phi_{+}=(2\Delta_{1}-d)\phi_{1}(x)-\phi_{2}(x)$ and $\phi_{-}=\phi_{1}(x)-(2\Delta_{2}-d)\phi_{2}(x)$ instead of $\phi_{1}(x)$ and $\phi_{2}(x)$. For each of $\phi_{+}=(2\Delta_{1}-d)\phi_{1}(x)-\phi_{2}(x)$ and $\phi_{-}=\phi_{1}(x)-(2\Delta_{2}-d)\phi_{2}(x)$ we have a conserved stress-energy tensor $T^{\pm}_{\mu\nu}$. In a quantum gravitational theory on this folded AdS$_{d+1}$, we should define a diffeomorphism (both large and small diffeomorphisms) invariant path integral measure with respect to the diffeomorphisms generated by the sum of these two stress-energy tensors $T^{+}_{\mu\nu}+T^{-}_{\mu\nu}$.

This measure is given by
\begin{equation}
\int D\phi_{+}D\phi_{-}e^{-i\int d^{d+1}x\sqrt{-g}\frac{1}{(2\Delta_{1}-d)^2+1}\Big(\phi_{+}(x)^2+\phi_{-}(x)^2\Big)}=1\,.\label{eq:measuregood}
\end{equation}
This can be used to fix the overall factor $C(\epsilon)$ of the path integral measure as following
\begin{equation}
\begin{split}
   & \int D\phi_{+}D\phi_{-}e^{-i\int d^{d+1}x\sqrt{-g}\frac{1}{(2\Delta_{1}-d)^2+1}\Big(\phi_{+}(x)^2+\phi_{-}(x)^2\Big)}\\=&\int C(\epsilon)\Pi_{\omega,\vec{k}}da_{0,\omega,\vec{k}}db_{0,\omega,\vec{k}}\Pi_{\lambda>0}da_{\lambda,\omega,\vec{k}}db_{\lambda,\omega,\vec{k}}\times\\&e^{-\int_{\omega^2-\vec{k}^2\geq0} \frac{d^{d-1}\vec{k}}{(2\pi)}d\omega\Big(\frac{(2\Delta_{1}-d-\frac{1}{2\Delta_{2}-d})^2}{(2\Delta_{1}-d)^2+1}(|a_{0,\omega,\vec{k}}|^2+\epsilon^{d-2\Delta_{1}}|b_{0,\omega,\vec{k}}|^2)+\sum_{\lambda>0}(|a_{\lambda, \omega,\vec{k}}|^2+|b_{\lambda, \omega,\vec{k}}|^2)\Big)}\\=&1\,,\label{eq:regu}
\end{split}
\end{equation}
where $\epsilon$ is the near asymptotic boundary cutoff $z=\epsilon\rightarrow0$ and we have used the fact that the modes $\delta\phi^{-}(x)_{\omega,\vec{k}}$ cannot be normalized to an $\epsilon$ independent constant under $\int d^{d+1}x\sqrt{-g}\phi^2(x)$ due to its asymptotic behavior $\delta\phi^{-}(x)_{\omega,\vec{k}}\sim z^{d-\Delta_{1}}$ (remember we have assumed that $\Delta_{1}>\frac{d}{2}$). Solving Equ.~(\ref{eq:regu}) we can see that the normalization constant $C(\epsilon)$ is a monomial of the cutoff scale $\epsilon$ which can heuristically be thought of as a Pauli-Villars regularization by which we can think of $C(\epsilon)$ as coming from integrating out free heavy fermions. This overall coefficient $C(\epsilon)$ will decouple in generating functionals for connected correlators. 

It is straightforward to see that the measure Equ.~(\ref{eq:measuregood}) is invariant under the diffeomorphisms 
\begin{equation}
x\rightarrow x, \quad g_{\mu\nu}(x)\rightarrow g'_{\mu\nu}(x)=\frac{\partial x'^{\rho}}{\partial x^{\mu}}\frac{\partial x'^{\sigma}}{\partial x^{\nu}}g_{\rho\sigma}(x')\,,\quad\phi_{+}(x)\rightarrow\phi'_{+}(x)=\phi_{+}(x')\,,\quad\phi_{-}(x)\rightarrow\phi'_{-}(x)=\phi_{-}(x')\,,
\end{equation}
which are indeed the diffeomorphisms generated by $T^{+}_{\mu\nu}+T^{-}_{\mu\nu}$.\footnote{Note that the measure Equ.~(\ref{eq:measuregood}) is also generally covariant under
\begin{equation}
x\rightarrow x', \quad g_{\mu\nu}(x)\rightarrow g'_{\mu\nu}(x')=\frac{\partial x^{\rho}}{\partial x'^{\mu}}\frac{\partial x^{\sigma}}{\partial x'^{\nu}}g_{\rho\sigma}(x(x'))\,,\quad\phi_{+}(x)\rightarrow\phi'_{+}(x')=\phi_{+}(x(x'))\,,\quad\phi_{-}(x)\rightarrow\phi'_{-}(x')=\phi_{-}(x(x'))\,.
\end{equation}}
However, it is interesting to notice that the measure Equ.~(\ref{eq:measuregood}) is not invariant under local diffeomorphisms generated by any other stress-energy tensors. For example it is not invariant under the diffeomorphisms generated by $T^{1}_{\mu\nu}(x)$ nor by $T^{2}_{\mu\nu}(x)$. This can be seen as following that in the folded AdS we have
\begin{equation}
\int d^{d+1}x\sqrt{-g}\frac{1}{(2\Delta_{1}-d)^2+1}\Big(\phi_{+}(x)^2+\phi_{-}(x)^2\Big)=\int d^{d+1}x \sqrt{-g}\Big(\phi_{1}(x)^2+\phi_{2}(x)^2\Big)\,,\label{eq:measurefinal}
\end{equation}
and just transforming $\phi_{1}(x)$ (to $\phi_{1}(x')$) or $\phi_{2}(x)$ (to $\phi_{2}(x')$) (with the corresponding transformation of the metric  $g_{\mu\nu}(x)\rightarrow g'_{\mu\nu}(x)=\frac{\partial x'^{\rho}}{\partial x^{\mu}}\frac{\partial x'^{\sigma}}{\partial x^{\nu}}g_{\rho\sigma}(x')$) will not leave it invariant.\footnote{Here however we notice that Equ.~(\ref{eq:measurefinal}) is invariant under the isometries generated solely by $T^{1}_{\mu\nu}(x)$ or $T^{2}_{\mu\nu}(x)$ as for them we have $d^{d+1}x'\sqrt{g(x')}=d^{d+1}x\sqrt{g(x)}$. This is important for the application of the AdS/CFT correspondence in the later analysis.}
This suggests that there are diffeomorphism anomalies associated with these broken diffeomorphisms and we cannot gauge these diffeomorphisms by coupling them with massless gravitons unless the anamoly is cancelled! 

We notice that there is a more elementary argument for the existence of such an anomaly if we consider quantum gravity in the original AdS$_{d+1}$. In this case we are gauging the symmetry generated by $T^{1}_{\mu\nu}(x)$. As a result, we should have the global part of this symmetry (the isometry) as symmetries in this quantum system. This tells us that we should be able to write down the symmetry generators for these global symmetries. However, the leakiness of $T^{1}_{\mu\nu}$ at the asymptotic boundary obstructs us to write down the generators as conserved charges
\begin{equation}
Q^{1}_{\zeta}=\int_{\Sigma}\zeta^{\mu}T^{1}_{\mu\nu} d\Sigma^{\nu}\,,\label{eq:charge}
\end{equation}
where $\Sigma$ is a Cauchy surface and $\zeta^{\mu}$ is the corresponding Killing vector. Hence we must have something to cancel the boundary flux such that charges Equ.~(\ref{eq:charge}) are conserved. The only possibility is that $T^{1}_{\mu\nu}$ is no longer divergence free which means that it is anomalous.
Nevertheless so far we don't know how to explicitly compute these diffeomorphism anomalies.\footnote{We should notice that standard results of  anomalies are that they are constrained by the Wess-Zumino consistency conditions and the solutions of these consistency conditions only exist in even dimensions \cite{Bardeen:1984pm}. More precisely gravitational anomalies was found to only exist $4n+2$ spacetime dimensions. However, these results only considered general covariance and transformation properties of the dynamical gauge field and they didn't consider the possible existence of background structures. The background structures such as the background metric are important for us since we are doing perturbative quantum gravity around a fixed background. 
}

Interestingly, this doesn't obstruct us to understand the physical implications of these diffeomorphism anomalies. To do so we notice that in the folded AdS$_{d+1}$ there would be two gravitons and they could form different linear combinations so this is a problem of matrix algebra. Let's focus on that graviton (i.e. a specific component of a matrix of two gravitons) which only couples to $\phi_{1}$. This is the graviton in the original AdS$_{d+1}$ universe in the unfolded picture where we couple the double-trace deformed scalar theory to dynamical gravity and we treat the graviton perturbatively. In this sector the path integral is given by
\begin{equation}
 Z=\int D[\phi_{1}]_{g}D[h_{\mu\nu}] e^{iS[\phi_{1},h;g]}\,,\label{eq:Z}
\end{equation}
where 
$h_{\mu\nu}$ is the perturbative graviton modes that is only coupled to $T^{1}_{\mu\nu}(x)$ and we emphasize that the path integral measure of $\phi_{1}$ depends on the metric $g_{\mu\nu}(x)$ which couples to both $\phi_{1}(x)$ and $\phi_{2}(x)$. The full metric on the original AdS$_{d+1}$ universe is given by
\begin{equation}
g_{\mu\nu}^{\text{full}}=g_{\mu\nu}+\sqrt{16\pi G_{N}}h_{\mu\nu}\,,
\end{equation}
where we treat $h_{\mu\nu}$ perturbatively and for our purpose we can think of $g_{\mu\nu}$ as the background metric.\footnote{Interestingly, this is the standard treatment in bimetric theories \cite{Damour:2002ws,Arkani-Hamed:2002bjr,Babichev:2009us,Babichev:2013usa}.} The diffeomorphism transform for the theory Equ.~(\ref{eq:Z}) is as following
\begin{equation}
\begin{split}
&x\rightarrow x, \quad g_{\mu\nu}(x)\rightarrow g'_{\mu\nu}(x)=\frac{\partial x'^{\rho}}{\partial x^{\mu}}\frac{\partial x'^{\sigma}}{\partial x^{\nu}}g_{\rho\sigma}(x')\,,\quad\phi_{1}(x)\rightarrow\phi'_{1}(x)=\phi_{1}(x')\,,\\&h_{\mu\nu}(x)\rightarrow h_{\mu\nu}'(x)=\frac{\partial x'^{\rho}}{\partial x^{\mu}}\frac{\partial x'^{\sigma}}{\partial x^{\nu}}h_{\rho\sigma}(x')+\nabla_{\mu}\epsilon_{\nu}(x)+\nabla_{\nu}\epsilon_{\mu}(x)\,,\quad \text{where } x'^{\mu}=x^{\mu}+\sqrt{16\pi G_{N}}\epsilon^{\mu}(x)\,,\label{eq:diffeotr}
\end{split}
\end{equation}
which works to the leading order in $G_{N}$ as the gauge transform in the perturbative treatment of graviton \cite{vanDam:1970vg,Donoghue:1994dn}. It transforms the action $S_{1}[\phi_{1},h;g]$ as following
\begin{equation}
\begin{split}
S_{1}[\phi_{1},h;g]\rightarrow S_{1}[\phi_{1}',h';g']&=S_{1}[\phi_{1},h^{\epsilon};g]\\&=S_{1}[\phi_{1},h;g]+2\sqrt{16\pi G_{N}}\int d^{d+1}x\sqrt{-g}\nabla^{\mu}\epsilon^{\nu}(x)T^{1}_{\mu\nu}(x)\,\\&=S_{1}[\phi_{1},h;g]-2\sqrt{16\pi G_{N}}\int d^{d+1}x\sqrt{-g}\epsilon^{\mu}(x)\nabla^{\nu}T^{1}_{\mu\nu}(x)\,,
\end{split}
\end{equation}
where in the last step we integrated by parts and used the fact that we consider $\epsilon^{\mu}(x)$ as a small diffeomorphism transform.\footnote{To be precise we consider $\epsilon^{\mu}\sim\mathcal{O}(z^{2})$ as $z\rightarrow0$ which is due to the fact that with our open boundary condition $\sqrt{-g}T_{\mu}^{1\nu}\sim \mathcal{O}(z^{-1})$ as $z\rightarrow0$ (which is easily obtained using the asymptotic behavior of $\phi_{1}(x)$ in Equ.~(\ref{eq:bcgr})).} Nevertheless from the above analysis of the path integral measure, there is a diffeomorphism anomaly 
\begin{equation}
\langle\nabla^{\mu}T^{1}_{\mu\nu}\rangle\,,\label{eq:anomaly}
\end{equation}
which obstructs the diffeomorphism invariance of the theory Equ.~(\ref{eq:Z}) at the quantum level.\footnote{Here we notice that the diffeomorphism anomaly Equ.~(\ref{eq:anomaly}) is a function of the background geometry $g_{\mu\nu}$ and graviton fluctuation $h_{\mu\nu}$. However, we don't need its explicit form for our later study.} This can be seen as following 
in the computation of the partition function Equ.~(\ref{eq:Z})
\begin{equation}
    \begin{split}
        Z&=\int D[\phi_{1}]_{g} D[h_{\mu\nu}]e^{iS[\phi_{1},h;g]}\,\\&=\int D[\phi_{1}']_{g'}D[h'_{\mu\nu}]e^{iS[\phi_{1}',h';g']}\,\\&=\int D[\phi_{1}']_{g'}D[h_{\mu\nu}]e^{iS[\phi_{1},h^{\epsilon};g]}\\&=\int D[\phi_{1}]_{g}e^{2i\sqrt{16\pi G_{N}}\int d^{d+1}x\sqrt{g}\epsilon^{\mu}\nabla^{\nu}\langle T^{1}_{\mu\nu}(x)\rangle}D[h_{\mu\nu}]e^{iS[\phi_{1},h^{\epsilon};g]}\,\\&=\int D[\phi_{1}]_{g}e^{2i\sqrt{16\pi G_{N}}\int d^{d+1}x\sqrt{g}\epsilon^{\mu}\nabla^{\nu}\langle T^{1}_{\mu\nu}(x)\rangle}D[h_{\mu\nu}]e^{iS[\phi_{1},h;g]-2i\sqrt{16\pi G_{N}}\int d^{d+1}x\sqrt{-g}\epsilon^{\mu}(x)\nabla^{\nu}T^{1}_{\mu\nu}(x)},\label{eq:gen}
    \end{split}
\end{equation}
where in the second step we used general covariance and in the fourth step we used the fact that the path integral measure is anomalous and we denoted the anomaly polynomial by $\langle\nabla^{\nu}T^{1}_{\mu\nu}(x)\rangle$. Hence we get the usual anomalous conservation law
\begin{equation}
\nabla^{\mu} T^{1}_{\mu\nu}(x)=\nabla^{\mu}\langle T^{1}_{\mu\nu}(x)\rangle\,,
\end{equation}
which gives us an operator identity that the divergence of the stress-energy tensor operators equals to the anomaly polynomial for the diffeomorphism transform.
\subsection{Non-Gravitational Bath}
The situation in the non-gravitational bath is similar to the case of gravitational bath. For simplicity we can take the geometry of the nongravitational bath also to be AdS$_{d+1}$ i.e. we turn off gravity in the gravitational bath. This can be done by sending the (dimensionless) Newton's constant $G_{N}^{\text{bath}}$ to zero (remember that we have set $l_{AdS}=1$). The analysis of the scalar fields is the same as the gravitational bath case. Hence the result is again that the diffeomorphism is anomalous in the gravitational AdS$_{d+1}$ and the treatment would be the same as Equ.~(\ref{eq:gen}).

\subsection{Restoring the Diffeomorphism Invariance via the St\"{u}ckelberg Mechanism}
Since we are considering a gravitational theory in the original AdS$_{d+1}$ universe, the diffeomorphism anomaly will potentially introduce several pathologies to the theory such as the breakdown of unitarity by inducing negative norm states. Hence, in a consistent description of the theory this anomaly should be cancelled. It can be cancelled 
if we introduce a vector field $V_{\mu}$ that transforms under the diffeomorphism as follows
\begin{equation}
 V_{\mu}(x)\rightarrow V_{\mu}'(x)=\frac{\partial x'^{\nu}}{\partial x_{\mu}}V_{\nu}(x')+\epsilon_{\mu}(x)\,,\quad\text{where}\quad x_{\mu}'=x_{\mu}+\sqrt{16\pi G_{N}}\epsilon_{\mu}(x)\,,
\end{equation}
and we couple it to the anomaly polynomial such that the anomaly can be compensated. This compensation works in the following theory
\begin{equation}
   Z_{\text{full}}=\int D[\phi_{1}]_{g}D[h_{\mu\nu}]D[V^{\rho}]e^{iS[\phi_{1},h;g]-2i\int d^{d+1}x\sqrt{-g}\sqrt{16\pi G_{N}}V_{\nu}(x)\langle\nabla^{\mu}T^{1}_{\mu\nu}(x)\rangle}\,,\label{eq:Zfull}
\end{equation}
where Equ.~(\ref{eq:Z}) can be thought of the gauge-fixed version of Equ.~(\ref{eq:Zfull}).\footnote{To get Equ.~(\ref{eq:Z}) from Equ.~(\ref{eq:Zfull}), the diffeomorphism gauge is completely fixed by setting $V^{\mu}(x)=0$. The fact that the theory is ghost free manifests only at the loop level via a Higgs mechanism (see later discussions).} The diffeomorphism invariance of Equ.~(\ref{eq:Zfull}) can be seen by either a direct diffeomorphism transformation or by thinking of $V^{\mu}(x)$ as a Lagrange multiplier which trivializes the diffeomorphism anomaly. Here we notice that for a theory without the diffeomorphism anomaly Equ.~(\ref{eq:anomaly}) is zero and Equ.~(\ref{eq:Zfull}) trivially equals to Equ.~(\ref{eq:Z}) upto a constant factor from the path integral of the vector field. Hence we can use Equ.~(\ref{eq:Zfull}) to extract the physical effects of the diffeomorphism anomaly if it is nonvanishing (more precisely nonvanishing in the matter sector).

\begin{figure}
\centering
\begin{tikzpicture}
      \draw[-] 
       decorate[decoration={zigzag,pre=lineto,pre length=5pt,post=lineto,post length=5pt}] {(-0.5,0) to (2.5,0)};
       \draw[-] 
       decorate[decoration={zigzag,pre=lineto,pre length=5pt,post=lineto,post length=5pt}] {(-0.5,0.2) to (2.5,0.2)};
       \draw[-, very thick] (2.5,0) arc (-180:180:1.5);
        \draw[-] 
       decorate[decoration={zigzag,pre=lineto,pre length=5pt,post=lineto,post length=5pt}] {(5.5,0) to (8.5,0)};
       \draw[-] 
       decorate[decoration={zigzag,pre=lineto,pre length=5pt,post=lineto,post length=5pt}] {(5.5,0.2) to (8.5,0.2)};
       \node at (-0.7,0.4)
       {\textcolor{black}{$\mu\nu$}};
       \node at (8.7,0.4)
       {\textcolor{black}{$\rho\sigma$}};
\end{tikzpicture}
\caption{The Feynman diagram which computes the effective action Equ.~(\ref{eq:S1}) from the matter loop which is a loop intergral of the correlator Equ.~(\ref{eq:TT}).} \label{pic:Feynman}
\end{figure}
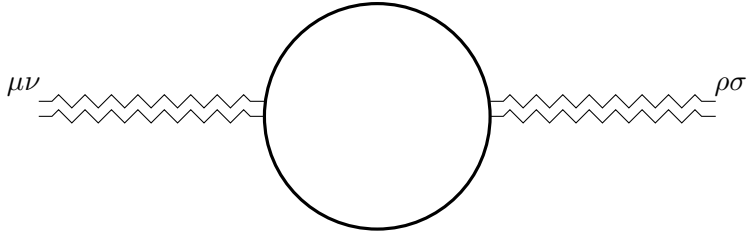

In fact we will show that graviton is massive in the fully diffeomorphism invariant theory Equ.~(\ref{eq:Zfull}). The graviton obtains the mass from a St\"{u}ckelberg mechanism.  This can be understood by again studying the partition function Equ.~(\ref{eq:Zfull}) and we do a diffeomorphism transform in the matter-graviton sector Equ.~(\ref{eq:diffeotr}) with respect to $x_{\mu}'=x_{\mu}+\sqrt{16\pi G_{N}}V_{\mu}(x)$,
\begin{equation}
\begin{split}
 Z_{\text{full}}&=\int D[\phi_{1}']_{g'}D[h'_{\mu\nu}]D[V^{\rho}]e^{iS[\phi_{1}',h';g']-2i\int d^{d+1}x\sqrt{-g}\sqrt{16\pi G_{N}}V^{\mu}(x)\langle\nabla^{\nu}T^{1}_{\mu\nu}(x)\rangle}\,\\&=\int D[\phi_{1}']_{g'}D[h_{\mu\nu}]D[V^{\rho}]e^{iS[\phi_{1},h^{V};g]-2i\int d^{d+1}x\sqrt{-g}\sqrt{16\pi G_{N}}V^{\mu}(x)\langle\nabla^{\nu}T^{1}_{\mu\nu}(x)\rangle}\,\\&=\int D[\phi_{1}]_{g}D[h_{\mu\nu}]D[V^{\rho}]e^{iS[\phi_{1},h^{V};g]}\,\\&=\int D[\phi_{1}]_{g}D[h_{\mu\nu}]D[V^{\rho}]e^{iS[\phi_{1},h;g]+2i\int d^{d+1}x\sqrt{-g}\sqrt{16\pi G_{N}}\nabla^{\mu}V^{\nu}(x)T^{1}_{\mu\nu}(x)}\,.\label{eq:coreZ}
\end{split}
\end{equation} 
We notice that we have the following gauge symmetry
\begin{equation}
h_{\mu\nu}(x)\rightarrow h_{\mu\nu}(x)+\nabla_{\mu}\epsilon_{\nu}(x)+\nabla_{\nu}\epsilon_{\mu}(x)\,,\quad V_{\mu}(x)\rightarrow V_{\mu}(x)-\epsilon_{\mu}(x)\,,\label{eq:new}
\end{equation}
in the resulting action in Equ.~(\ref{eq:coreZ}). Now we can integrate out the matter field to get an effective theory $S_{\text{eff}}[V,h;g]$ involving only the vector field $V^{\mu}(x)$ and graviton $h_{\mu\nu}(x)$ which should be invariant under the gauge symmetry Equ.~(\ref{eq:new}). A kinetic term 
\begin{equation}
 S_{1}\sim\int d^{d+1}x\sqrt{-g} \Big(\nabla_{\mu}V_{\nu}(x)\nabla^{\mu}V^{\nu}(x)+\nabla_{\mu}V_{\nu}(x)\nabla^{\nu}V^{\mu}(x)\Big)\,,\label{eq:S1}
\end{equation}
of the vector field will be generated by a matter loop which is described by the time-ordered two-point function of the matter stress-energy tensor (see Fig.\ref{pic:Feynman})
\begin{equation}
\langle T\Big(T^{1}_{\mu\nu}(x)T^{1}_{\rho\sigma}(y)\Big)\rangle\,.\label{eq:TT}
\end{equation}
However, the invariance under Equ.~(\ref{eq:new}) obligates that the full effective action $S_{\text{eff}}[V,h]$ should be a function of the Equ.~(\ref{eq:new}) invariant combination
\begin{equation}
\nabla_{(\mu}V_{\nu)}+h_{\mu\nu}\equiv\nabla_{\mu}V_{\nu}+\nabla_{\nu}V_{\mu}+h_{\mu\nu} \,.
\end{equation}
Hence the quadratic action Equ.~(\ref{eq:S1}) should in fact be
\begin{equation}
 S_{1}'\sim \int d^{d+1}x \sqrt{-g}\Big(\nabla_{(\mu}V_{\nu)}+h_{\mu\nu}\Big)\Big(\nabla^{(\mu}V^{\nu)}+h^{\mu\nu}\Big)\,,
\end{equation}
which indeed is a St\"uckelberg mass term of the graviton $h_{\mu\nu}$ and the exact value of the mass is given by the matter loop described by the two-point function Equ.~(\ref{eq:TT}). This is precisely the result of \cite{Porrati:2001db}. Moreover, similar to \cite{Preskill:1990fr} the manifestly diffeomorphism invariant description Equ.~(\ref{eq:Zfull}) should be able to be easily uplifted to a full description as its Higgs phase where the diffeomorphism is spontaneously broken and $V^{\mu}(x)$ is the Goldstone vector boson. More precisely, to extract the exact value of the graviton mass we have to extract the delta-function piece of the two-point function Equ.~(\ref{eq:TT}). We provide an example of such calculation in the next subsection.

\subsection{An Example of Graviton Mass Extraction}\label{sec:demon}
In this subsection, we will consider a specific example to demonstrate the algorithm of graviton mass extraction we proposed in the previous subsection. We will consider a conformally coupled scalar field in four-dimension anti-de Sitter space.\footnote{See \cite{Geng:2023ta1} for a more efficient way to extract the graviton mass for generic situations.} This means that $m^{2}=\frac{d-1}{4d}R$ where $R$ is the Ricci scalar which takes the value $-d(d+1)$ and $d+1=4$. We can check that $\Delta_{-}=\frac{d}{2}-\frac{1}{2}\sqrt{d^2+4m^2}=\frac{d-1}{2}>\frac{d-2}{2}$ so this is a good example fitting in the formalism we developed above. From now on we take $d=3$.

We will use the embedding space formalism. In this formalism, the Lorentzian AdS$_{4}$ is a submanifold
\begin{equation}
-X_{0}^2-X_{4}^2+X_{1}^2+X_{2}^2+X_{3}^2=-1\,,\label{eq:embedding}
\end{equation}
of an ambient five-dimensional two-time Minkowski spacetime which has the metric
\begin{equation}
ds^{2}=-dX_{0}^2-dX_{4}^2+dX_{1}^2+dX_{2}^2+dX_{3}^2\,.
\end{equation}
The Poincar\'{e} patch is given by the following parametrization of Equ.~(\ref{eq:embedding})
\begin{equation}
\begin{split}
X_{0}&=\frac{z^2+x_{1}^2+x_{2}^2-t^2+1}{2z}\,,\\
X_{1}&=\frac{x_{1}}{z}\,,\\
X_{2}&=\frac{x_{2}}{z}\,,\\
X_{3}&=\frac{z^2+x_{1}^2+x_{2}^2-t^2-1}{2z}\,,\\
X_{4}&=\frac{t}{z}\,,
\end{split}
\end{equation}
and the Euclidean AdS$_4$ is obtained by 
\begin{equation}
    X_{4}\rightarrow iX_{4}\,,\quad t\rightarrow it\,.
\end{equation}
Moreover, all the fields in AdS$_{4}$ are uplifted homogeneously to fields in the ambient space and these ambient space fields reduce to the AdS$_{4}$ fields when restricted on the submanifold Equ.~(\ref{eq:embedding}). The homegeneity conditions are
\begin{equation}
X^{M} S_{MNPQ\cdots}(X)=0\,\quad \text{and} X^{L}\nabla_{L}S_{MNPQ\cdots}(X)=nS_{MNPQ\cdots}\,,\label{eq:emb} 
\end{equation}
where $X^{M}$ denotes the coordinate in the ambient space, 
 $n$ is called the degree and $S_{MNPQ\cdots}$ refers to the uplift of a generic AdS$_{4}$ tensorial field $S_{\mu\nu\rho\sigma\cdots}$ (the rank of the tensor doesn't change before and after the uplift). The most relevant property to us is that a divergenceless AdS tensor is uplifted to a divergenceless tensor in the ambient space. For details we refer the readers to \cite{Porrati:2001db,Duff:2004wh,Aharony:2006hz, Apolo:2012gg} and references therein.

We will follow the formalism in \cite{Duff:2004wh} and quote their results. For the sake of simplicity we go to the Euclidean coordinates. Our purpose is to extract the graviton mass from the following integral
\begin{equation}
    \int_{AdS_{4}} dY \langle  T^{1}_{MN}(x)T^{1}_{PQ}(y)\rangle\,.
\end{equation}
Remember that we can do this because the effective action of the emergent Goldston vector field is
\begin{equation}
   S_{1}= -\frac{4}{2}16\pi G\int_{AdS_{d+1}} dX dY \nabla^{M}V^{N}(X)\langle T^{1}_{MN}(X)T^{1}_{PQ}(Y)\rangle\nabla^{P}V^{Q}(Y),\label{eq:target}
\end{equation}
and we are interested in the fintie short distance term generated from it which is of the form
\begin{equation}
 \mathcal{L}_{1}=-\frac{m^2}{2}\nabla_{(M}V_{N)}(x)\nabla^{(M}V^{N)}(x)\,,
\end{equation}
where $m^2$ is the graviton mass due to the St\"uckelberg mechanism. Since we are considering a conformally coupled scalar field in AdS$_4$ whose stress-energy tensor is traceless, we only have to consider a divergenceless vector field $V^{\mu}(x)$ in AdS$_4$ which is uplifted to a divergenceless vector $V^{M}(X)$ in the ambient space. Furthermore because $\langle T^{MN}(X)T^{PQ}(Y)\rangle$ is already transverse and traceless in $MN$ and $PQ$, we only need the contributions proportional to $\delta^{(M}_{(P}\delta^{N)}_{Q)}$ from it as other contributions vanish when inserting to Equ.~(\ref{eq:target}) in the limit $y\rightarrow x$.

We will use existing results from \cite{Duff:2004wh}. The basic result we will use is Equ.(35) from \cite{Duff:2004wh} and as we explained in the previous paragraph we only need the $\mathcal{O}_{3}=2\delta^{(M}_{(P}\delta^{N)}_{Q)}$ term\footnote{It may not be straightforward to see that $Y^{N}Y^{M}\nabla_{M}V_{N}(Y)=0$ so let's prove it here. Differentiating by parts we have $Y^{N}Y^{M}\nabla_{M}V_{N}(Y)=Y^{M}\nabla_{M}(Y^{N}V_{N})-Y^{N}V_{N}$ where the resulting two terms are both zero due to $Y^{N}V_{N}(Y)=0$. Moreover, the degree of the vector $V^{M}$ has to be $1$ in order for the consistency with diffeomorphism invariance. This is because the graviton $h_{MN}$ are traceless symmetric and satisfies the first equation in Equ.~(\ref{eq:emb}) which implies that only $\mathcal{O}_{3}$ in \cite{Duff:2004wh} is relevant to the 1-loop correction of graviton propagator. Hence this should also be the case for the vector field and in particular $\mathcal{O}_{5}$ in \cite{Duff:2004wh} shouldn't contribute which requires $n=1$ for $V^{M}$.} which is
\begin{equation}
    \frac{4}{9}\Delta_{0}'(Z)^2+\frac{1}{9}\Delta_{0}(Z)\Delta_{0}''(Z).
\end{equation}
Here $Z=-X\cdot Y$ is the invariant distance between the two points $x^{\mu}$ and $y^{\nu}$ in AdS$_{4}$ with $X^{M}$ and $Y^{M}$ their embedding space coordinates and $\Delta_{0}(Z)$ is the propagator given as Equ.(28) in \cite{Duff:2004wh}
\begin{equation}
\Delta_{0}(Z)=\frac{1}{8\pi^2 L^2}(\frac{a}{Z+1}+\frac{b}{Z-1})\,,
\end{equation}
where $a=\beta-\alpha$ and $b=\alpha+\beta$. Using the embedding space coordinates for the Poincare patch we can find that
\begin{equation}
Z=\frac{((\vec{x}-\vec{y})^2 + z_{1}^2 + z_{2}^2)}{2 z_{1} z_{2}}=\frac{r^2+z_{1}^2+z_{2}^2}{2z_{1}z_{2}}\,.
\end{equation}
Here we notice that for the conformal scalar field we have $\Delta_{+}=\frac{d+1}{2}=2$ and $\Delta_{-}=\frac{d-1}{2}=1$ and as we send $x^{\mu}$ close to the asymptotic boundary we have $z_{1}\rightarrow0$ and
\begin{equation}
\begin{split}
\Delta_{0}(Z)&\sim \frac{1}{4\pi^2 L^2}\Big[\beta (z_{1}+\mathcal{O}(z_{1}^{3}))+\alpha (z_{1}^2+\mathcal{O}(z_{1}^{4})\Big]\\&=\frac{1}{4\pi^2 L^2}\Big[\beta (z_{1}^{\Delta_{-}}+\mathcal{O}(z_{1}^{\Delta_{-}+2}))+\alpha (z_{1}^{\Delta_{+}}+\mathcal{O}(z_{1}^{\Delta_{+}+2})\Big]\,,\label{eq:asymres}
\end{split}
\end{equation}
which is consistent with our results of double-trace deformation Equ.~(\ref{eq:witten}), Equ.~(\ref{eq:bcnong}) and Equ.~(\ref{eq:bcgr}).

Now we can compute the following integral
\begin{equation}
\begin{split}
  \int_{AdS_4}dY\Big(\frac{4}{9}\Delta_{0}'(Z)^2+\frac{1}{9}\Delta_{0}(Z)\Delta_{0}''(Z)\Big)\,\\=\int_{0}^{\infty} \frac{dz_{1}}{z_{1}^4}\int_{0}^{\infty}4\pi r^2 dr\Big(\frac{4}{9}\Delta_{0}'(Z)^2+\frac{1}{9}\Delta_{0}(Z)\Delta_{0}''(Z)\Big),
\end{split}
\end{equation}
where we have to split the $z_{1}$ integral as $\int_{0}^{z_{2}-\epsilon}dz_{1}+\int_{z_{2}+\epsilon}^{\infty}dz_{1}$ and expand the result in the UV-cutoff $\epsilon$. The result is
\begin{equation}
\begin{split}
&\frac{2\pi^2 b^2 z_{2}^4}{3 \epsilon ^4}+\frac{2\pi^2  a b   z_{2}^2}{9 \epsilon ^2}+\frac{\pi^2}{9} \left(3a^{2}-3b^{2}-6 a b+16ab\log(\frac{z_{2}}{\epsilon^{2}}) \right)+\frac{\pi^2\epsilon a \left(74 b -3 a\right)}{36 z_{2}}+O\left(\epsilon ^2\right)\,,
\end{split}
\end{equation}
multiplied by $\frac{1}{(8\pi^2 L^2)^2}$. This result is of fundamental importance to us. Since we only want to extract the delta-function piece of the two-point function Equ.~(\ref{eq:TT}), we only need the $\mathcal{O}(1)$ terms.  The two power law divergent terms are universal short distance singularities which can be removed by an appropriate point splitting regularization. The terms proportional to $ab$ in the $\mathcal{O}(1)$ order are removed by appropriate counterterms which cures the logarithmic divergence. This is because $ab=\beta^2-\alpha^2$ and it is hence independent of having the double-trace deformation which tangles the two boundary conditions (see Equ.~(\ref{eq:asymres})). Thus the $ab$ term persists even when we are doing the standard quantization in AdS/CFT i.e. $\beta=0$. Hence the relevant term to us is
\begin{equation}
\frac{\pi^2}{3(8\pi^2L^2)^2}(a^2-b^2)\,,
\end{equation}
which in terms of $\alpha$ and $\beta$ is
\begin{equation}
-\frac{4\pi^2}{3(8\pi^2 L^2)^2}\alpha\beta.
\end{equation}
Now the graviton mass can be extracted as
\begin{equation}
m^2=4\times16\pi G\times 2\frac{4\pi^2}{3(8\pi^2L^2)^2}\alpha\beta=\frac{8G}{3\pi L^4}\alpha\beta\,,\label{eq:result}
\end{equation}
where the factor $2$ comes from the factor $2$ in the definition of $\mathcal{O}_{3}$ and the discrepancy of the coefficient with the results in \cite{Porrati:2001db,Duff:2004wh} is a matter of convention of how we normalize the vector field.\footnote{We note that even the results in \cite{Porrati:2001db} and \cite{Duff:2004wh} are different with the coefficients.} The physical relevance of this result is that the graviton mass is zero when either $\alpha$ or $\beta$ is zero which reduces to either the standard or alternative quantizations in the usual AdS/CFT and it is only nonzero when both of $\alpha$ and $\beta$ are nonzero which is the case that we have a bath glued.

\subsection{The Unbroken Diffeomorphism Invariance with a Gravitational Bath}\label{sec:gravbathml}
We have analyzed the diffeomorphism broken and graviton mass generation in the AdS$_{d+1}$ bulk when a bath is glued on its asymptotic boundary. This result persists for both cases when the bath is nongravitational and when it is gravitational. Nevertheless, in the case with a gravitational bath, we noticed that the diffeomorphism invariance generated by $T^{1}_{\mu\nu}+T^{2}_{\mu\nu}$ is non-anomalous.
This is consistent with the fact that the total stress-energy of the folded AdS$_{d+1}$ (or the union of the original two AdS$_{d+1}$'s) is conserved (this has a nice interpretation using the AdS/CFT see Sec.~\ref{sec:discc} for discussions on this point).

This result suggests that in this resulting universe from folding, apart from a massive graviton, there is also a massless graviton. This is the hinge of the recent realization that wedge holography provides a doubly holographic set-up for massless gravity localized on the Karch-Randall braneworld which can be used to demonstrate that there is no entanglement island in massless gravity theories \cite{Geng:2020fxl,Geng:2021hlu,Geng:2023qwm}.

\subsection{Discussions}\label{sec:discc}
In this section, we carefully analyzed the quantum effects encoded in the AdS$_{d+1}$ scalar field partition function. In particular, we noticed that we are not able to define a diffeomorphism invariant path integral measure 
in the AdS$_{d+1}$ universe that is glued to a bath by a marginal double-trace deformation. This suggests the possible existence of a diffeomoprhism anomaly in the AdS$_{d+1}$ universe. Although we don't know how to compute this anomaly explicitly to confirm its existence
, we found that we can extract its physical effect by coupling the theory with dynamical gravity in a diffeomorphism invariant way by introducing a compensating vector field to cancel the anomaly. We notice that if the diffeomorphism anomaly doesn't exist then this coupling is a trivial operation. However, our calculation shows that in the cases with double-trace deformation this anomaly should exist and it indicates a spontaneous breaking of the diffeomorphism anomaly. This is signaled by the graviton mass Equ.~(\ref{eq:result}) which is generated from a St\"{u}ckelberg mechanism. In fact we can show that the graviton will be massive even if the diffeomorphism anomaly turns out to be zero (see Appendix.\ref{sec:A}).

We can see from 
Equ.~(\ref{eq:result}) that the graviton is massive in both the double-trace deformation in standard AdS/CFT where $\alpha$ and $\beta$ are not independent (see Sec.~\ref{sec:dbt}) and the case with a bath glued where $\alpha$ and $\beta$ are independent (see Sec.~\ref{sec:dbto}). This result is consistent with the lore in AdS/CFT that the bulk graviton mass duals to the anomalous dimension of the stress-energy tensor of CFT$^{1}$. In both the standard quantization and alternative quantization cases this anomalous dimension is zero due to the conservation of the CFT$^{1}$ stress-energy tensor. Nevertheless, in the cases with a double-trace deformation the CFT$^{1}$ stress-energy tensor is no longer conserved due to this deformation
\begin{equation}
\partial_{a}T^{ab}_{\text{CFT$^{1}$}}\propto \partial^{b} (\text{double-trace deformation})\,.\label{eq:descendant}
\end{equation}
Hence its anomalous dimension is not protected to be zero anymore. Interestingly, the equation Equ.~(\ref{eq:descendant}) also suggests that the stress-energy tensor has a vector as its conformal descendant which through the AdS/CFT correspondence translates to AdS$_{d+1}$ as the statement that the graviton is no longer in a short multiplet of the AdS$_{d+1}$ isometry group $SO(d,2)$ but in a long-multiplet containing a vector field (remember we have shown that the AdS$_{d+1}$ isometry group is not anomalous in Sec.~\ref{sec:gravbathana}.). This is nothing but the holographic interpretation of the existence of the vector field $V^{\mu}(x)$ in our calculation. Fixing the gauge by $V^{\mu}(x)=0$ is a manifestation of the Higgs mechanism \cite{Porrati:2001db}. This further confirms that the AdS$_{d+1}$ diffeomorphism is spontaneously broken and the graviton becomes massive when there is a double-trace deformation. 

Nevertheless we should notice that there may exist quantum aspects in AdS$_{d+1}$, which are expected from the CFT side, that cannot be extracted by our calculation. An example is for the double-trace deformation in standard AdS/CFT that we studied in Sec.~\ref{sec:dbt}. In this case, in the dual CFT there indeed exists a conserved stress-energy tensor which is the stress-energy tensor of the undeformed CFT plus a contribution from the double-trace deformation. It is however not clear to us what is the dual statement in the AdS$_{d+1}$ which may be that there is another massless spin-2 modes that overlaps with the massive graviton. This expectation comes from the following consideration. In the gravitational bath case Sec.~\ref{sec:gravbath}, as opposed to the nongravitational bath case in Sec.~\ref{sec:nongravbath}, we can dualize the union of AdS$_{d+1}$ and the gravitaitonal bath to a d-dimensional conformal field theory which is CFT$^{1}$+CFT$^{2}$ and together with the double-trace deformation that couples them together (remember we've taken the double-trace deformation to be marginal). Hence from the CFT perspective we do have a conserved d-dimensional CFT stress-energy tensor and this should be dual to a massless spin-2 mode that propagates in both the AdS$_{d+1}$ and the bath. It is shown in \cite{Aharony:2006hz} that this massless mode is a specific linear combination of the gravitons from the AdS$_{d+1}$ and the bath. This is consistent with our analysis in Sec.~\ref{sec:gravbathana} and Sec.~\ref{sec:gravbathml} that we can indeed define a diffeomorphism invariant measure for the matter fields in the folded AdS$_{d+1}$ for the diffeomorphisms generated by $T^{1}_{\mu\nu}+T^{2}_{\mu\nu}$. Nevertheless the calculation in Sec.~\ref{sec:demon} still suggests that there is another linear combination of the two graviton modes which is massive and is orthogonal to the massless one. Hence the graviton in the original AdS$_{d+1}$ is a superposition of the massive and the massless modes and hence overlaps with both of them. Therefore we expect that in the case of the double-trace deformation in standard AdS/CFT we should have a similar result for the consistency with the CFT expectation. However, it is possible that the actual situation is subtler than our expectation. For example the double-trace deformation Equ.~(\ref{eq:dtdef}) in standard AdS/CFT is not marginal (remember we have chosen $\Delta=\frac{d}{2}+\frac{1}{2}\sqrt{d^2+4m^2}$ such that we can perform the alternative quantization $\frac{d+2}{2}>\Delta>\frac{d}{2}$ otherwise the modes Equ.~(\ref{eq:dtmode}) wouldn't be normalizable in the Klein-Gordon measure) so the theory with the deformation Equ.~(\ref{eq:dtdef}) wouldn't be stable under the RG flow and the correct low energy behavior should be captured by either the limit $h\rightarrow0$ or $h\rightarrow\infty$ depending on whether the deformation Equ.~(\ref{eq:dtdef}) is irrelevant or relevant. Indeed, in this two limits we are in the standard AdS/CFT without any deformation (i.e. either $\alpha$ or $\beta$ in Equ.~(\ref{eq:result}) is zero\footnote{Notice that on the one hand if we did the standard quantization before we turn on the double-trace deformation we have $\alpha=\frac{}8h{d-2\Delta_{+}}\Big(\pi^{\frac{d}{2}}\frac{\Gamma[\Delta_{+}+1-\frac{d}{2}]}{\Gamma[\Delta_{+}]}\Big)^{2}\beta$ (see Equ.~(\ref{eq:ab1})). In this case the double-trace deformation is irrelevant so $h\rightarrow0$ is the correct limit for low energy physics and we have to fix a finite $\beta$ then we have $\alpha=0$. On the other hand, if we start with the alternative quantization then we have $\alpha=-\frac{4h}{d-2\Delta_{-}}\Big(\pi^{\frac{d}{2}}\frac{\Gamma[\Delta_{-}+1-\frac{d}{2}]}{\Gamma[\Delta_{-}]}\Big)^{2}\beta$. In this case the double-trace deformation is relevant so $h\rightarrow\infty$ is the correct limit for low energy physics and we have to fix $\alpha$ to be a finite number then we have $\beta=0$.\label{ft:12}}) and the graviton is fact massless so we only end up with a single massless graviton. We leave this question to future work.

\section{Gravitational Subregion and Edge Modes}\label{sec:subr}
The model with a gravitational bath glued on has another interesting application. This is from the observation that the original AdS$_{d+1}$ universe can be thought of as a subregion of a gravitational universe which consists of the union of the original AdS$_{d+1}$ and the bath AdS$_{d+1}$. Moreover, this subregion is by itself gravitational and the nice feature of it is that the gravity turns off kinematically near its boundary (the asymptotic boundary). Hence this provides a tractable model of a gravitational subregion without the necessity to address the issue of the graviton edge modes \cite{Donnelly:2016auv}. As a result, we can study the entanglement entropy between a gravitational subregion and its complement in this model (see \cite{Geng:2020fxl} for a holographic study of this question using the Karch-Randall braneworld). 

Nevertheless, we do have an edge mode to deal with in the matter sector. As we split the AdS$_{d+1}$ and the bath by integrating out the bath and focus on the AdS$_{d+1}$, the source $J(x)$ (see Sec.~\ref{sec:partitons}) becomes the edge mode which we have to integrate over in the resulting path integral. Hence we have
\begin{equation}
\begin{split}
Z_{AdS_{d+1}}&=\det(\Box-m^2)^{-\frac{1}{2}}\int DJ Z[J]_{\text{CFT}}\,\\&= \det(\Box-m^2)^{-\frac{1}{2}}\int DJ e^{i\pi^{\frac{d}{2}}\frac{\Gamma[\Delta_{+}+1-\frac{d}{2}]}{\Gamma[\Delta_{+}]}\int dtd^{d-1}\vec{x}dt'd^{d-1}\vec{x'}\frac{J(t',\vec{x'})J(t,\vec{x})}{(-(t-t')^2+|\vec{x}-\vec{x'}|^{2})^{\Delta_{+}}}}\,,
\end{split}
\end{equation}
where $Z[J]_{\text{CFT}}$ is given by Equ.~(\ref{eq:treeZs}) and the $\det(\Box-m^2)^{-\frac{1}{2}}$ is the quantum correction. We can simplify the expression by noticing that in $d$-dimensions (i.e. the boundary of the AdS$_{d+1}$) we have
\begin{equation}
\nabla^2 \frac{1}{\Big(-(t-t')^2+|\vec{x}-\vec{x}'|^2\Big)^{\frac{d-2}{2}}}=\delta(t-t')\delta^{d-1}(\vec{x}-\vec{x'})\,.
\end{equation}
So we get
\begin{equation}
\begin{split}
Z_{AdS_{d+1}}&= \det(\Box-m^2)^{-\frac{1}{2}}\int DJ e^{i\pi^{\frac{d}{2}}\frac{\Gamma[\Delta_{+}+1-\frac{d}{2}]}{\Gamma[\Delta_{+}]}\int dtd^{d-1}\vec{x}dt'd^{d-1}\vec{x'}\frac{J(t',\vec{x'})J(t,\vec{x})}{(-(t-t')^2+|\vec{x}-\vec{x'}|^{2})^{\Delta_{+}}}}\,\\&=\det(\Box-m^2)^{-\frac{1}{2}}\det(\nabla^2)^{\frac{2\Delta_{+}}{d-2}}\,,
\end{split}
\end{equation}
where the second factor is from the edge mode $J(t,\vec{x})$ and it of positive power of the determinant of the $d$-dimensional d'Albembert. This positive power is of a wrong sign for bosonic fields and is exactly the expected property for edge modes \cite{Donnelly:2014fua}.
\section{Conclusion and Future Directions}\label{sec:conclusion}
In this paper, we carefully studied the AdS/CFT model of coupling a gravitational universe to a bath at its asymptotic boundary. The bath is modelling an experimental laboratory that is trying to extract the physics in the gravitational universe. This model avoids the complicated question of describing or modelling an observer in a gravitational universe and hence it is of direct relevance to build the bridge between quantum gravity and experiment. 

In this AdS/CFT model the coupling is achieved by a double-trace deformation that in the CFT description couples the AdS-dual CFT to the bath. We carefully analyzed the quantum aspects of this model by doing canonical quantization and studying the partition function at the quantum level. We notice that the bulk Hilbert space is enlarged due to the bath coupling and the double-trace deformation generically induces a mass for the graviton in the gravitating universe through a St\"{u}ckelberg mechanism. In other words, the diffeomorphism symmetry in the gravitational universe is spontaneously broken if it is coupled to a bath. This suggests that we may have to think of quantum gravity quite differently from a practical point of view.

Moreover, this work is potentially helpful to get a proper understanding of subregion physics in a gravitational universe. In the case of the (Yang-Mills) gauge theory, a subregion can be defined gauge invariantly by introducing edge modes which splits the manifold into two regions $\mathcal{R}$ and $\bar{\mathcal{R}}$ and assigns opposite surface charges to $\partial\mathcal{R}$ and $\partial\bar{\mathcal{R}}$ \cite{Donnelly:2014fua}. Nevertheless, in the case of a gravitational universe with a massless graviton we can't have negative mass density and therefore the idea of edge modes wouldn't work in the same way as in the case of gauge theory.\footnote{Though in the low dimensional case where the gravitational theory is topological it might be possible to do this by extending the physical Hilbert space and do a careful projection in the end (see \cite{Lin:2018xkj,Jafferis:2019wkd,Mertens:2022ujr,Wong:2022eiu,Chua:2023ios} for recent attempts).} Hence if we constrain ourselves to a subregion the subregion diffeomorphism invariance will be broken and the question is how to restore it and what is the consequences of the restoration mechanism. This paper provides a mechanism by introducing a St\"{u}ckelberg field which compensates the diffeomorphism anomaly. This naively renders the subregion graviton to be massive. 
But as opposed to the situation we considered in this paper where there is a clear connection between diffeomorphism anomaly and the openness of the system which manifests using the folding trick, in the general subregion case we don't know if we still have the issue of the disability to define a diffeomorphism invariant path integral measure. If not the stress-energy tensor is locally conserved and free of anomaly. The issue is then that the stress-energy flux is nonzero on the boundary of the subregion $\mathcal{R}$ and so the diffeomorphism symmetry $\zeta^{\mu}$ for whom $\zeta^{\mu}n^{\nu}T_{\mu\nu}|_{\partial \mathcal{R}}\neq 0$ is broken. 
To restore such diffeomorphisms\footnote{We should emphasized that the motivation to restore this diffeomorphism symmetry is different from the case studied in the paper. This is motivated by the attempt to define a gravitaional subregion and studying the associated entanglement entropy. This diffeomorphism is a symmetry for the total universe $\mathcal{R}\cup\bar{\mathcal{R}}$ so if we integrate out the complementary region $\bar{\mathcal{R}}$ in the path integral the symmetry should persists in the resulting theory on $\mathcal{R}$. We should notice here that in gravity the situation is very different from the gauge theory case \cite{Donnelly:2014fua} where we could interpret the surface gauge symmetry on $\partial{\mathcal{R}}$ as assigning charges of opposite signs to $\partial{\mathcal{R}}$ and $\partial{\bar{\mathcal{R}}}$ and splitting the Hilbert space of $\mathcal{R}$ into different superselection sectors (associated with different values of the surface charge) as in gravity we cannot have negative surface mass.} it is enough to have a St\"{u}ckelberg field localized on the boundary $\partial\mathcal{R}$ which compensates the diffeomorphism transform of the action due to the nonzero flux. It's not clear at this stage what the implications of this St\"{u}ckelberg mechanism are. We leave this question to future work.

Another important question to study in the future is to properly understand the situation in the double-trace deformation for the standard AdS/CFT. As discussed at the end of Sec.~\ref{sec:discc}, it is interesting to know whether the graviton is massive in this situation. Here we further notice that we have $\alpha\propto h\beta$ in the double-trace deformation for the standard AdS/CFT where $\alpha$ and $\beta$ are asymptotic coefficients of the bulk Green function. This relationship is easily seen from Equ.~(\ref{eq:ab1}) and Equ.~(\ref{eq:ab2}). Hence we have from Equ.~(\ref{eq:result}) that the graviton mass square is linear in $h$ which suggests an instability for positive $h$ either we started with the standard or the alternative quantization when $h=0$. However, in the case that we coupled the original AdS$_{d+1}$ and the bath by a double-trace deformation we have independent $\alpha$ and $\beta$ and the graviton mass square is proportional to the square of the double-trace deformation parameter which signals a stable configuration. This can be seen from Equ.~(\ref{eq:canonicalgrav}) that the two sets of modes $\delta\phi(x)^{\pm}_{\omega,\vec{k}}$ are assigned with independent creation and annihilation operators and so the Green function would be the sum of their contributions and there is no mixing term between term. Using the fact that if we fix $\delta\phi^{+}(x)_{\omega,\vec{k}}$ to be independent of $g$ then $\delta\phi^{-}(x)^{-}_{\omega,\vec{k}}$ will be linear in $g$, we see that the Green function only contains terms scaling with $g$ as $g^{0}$ and $g^{2}$ (i.e. $\alpha\propto g^{2}$ and $\beta\propto g^{0}$). Hence the graviton mass square is proportional to $g^2$ (see \cite{Aharony:2006hz} for an explicit calculation). Thus there is no instability issue in this case with a bath glued. This analysis suggests that with the double-trace deformation in the standard AdS/CFT 
we should necessarily consider the bulk backreaction to the geometry if $h$ is finitely positive (but close enough to zero as we know when $h\rightarrow\infty$ we have $\beta=0$ and so zero graviton mass square (see footnote.\ref{ft:12})). 
It would be interesting to exploit these observations further.

More practically, our model can be used to study dynamics of strongly-coupled open quantum systems using both CFT and holographic techniques. For example, in an real experimental system disordering, dissipation and spatial inhomogeneity are unavoidable and may have important experimental and theoretical implications such as the Anderson localization \cite{PhysRev.109.1492} and dissipative effects in hydrodynamics, heavy ion collisions and quantum phase transitions. Coupling the system with a bath in a control manner as in our system provides an explicit model to study these effects. The holographic dual of our system provides a nice setup to explore these effects in strongly-coupled systems. We can also turn on a finite temperature in our model by considering the bulk to be a black hole geometry (and the bath also to be of a finite temperature). It would be interesting to carry out these analysis in detail using our holographic model (see \cite{Kiritsis:2011zq,Blake:2013bqa,Adams:2014vza,Huang:2023ihu} for early and recent studies of some of these questions). 
\section*{Acknowledgements}
We are grateful to Amr Ahmadain, Luis Apolo, Jan de Boer, Sylvain Fichet, A.~Liam Fitzpatrick, Eduardo Gonzalo, Brianna Grado-White, Indranil Halder, Matthew Heydeman, Yangrui Hu, Yikun Jiang, Juan Maldacena, Martin Sasieta, Brian Swingle, Gabriel Wong for relevant discussions. We thank Daniel Grumiller, Andreas Karch, Suvrat Raju, Lisa Randall, Romain Ruzziconi and C\'{e}line Zwikel for discussions and relevant collaborations on \cite{Grumiller:2023ahv}. HG would like to thank the organizers of ``Strings 2023" conference at the Perimeter Institute for Theoretical Physics where the final stage of this work is done. The work of HG is supported by the grant (272268) from the Moore Foundation ``Fundamental Physics from Astronomy and Cosmology'' and a grant from Physics Department at Harvard University. 

\appendix
\section{A Potential Loophole and Its Resolution}\label{sec:A}
We should notice that there is a potential loophole in our argument leading to Equ.~(\ref{eq:coreZ}). As we discussed, at this point we don't know how to compute the diffeomorphism anomaly $\nabla^{\mu}\langle T^{1}_{\mu\nu}(x)\rangle$. Hence it is reasonable to doubt its very existence. However, in this appendix we will show that even if the diffeomorphism anomaly $\nabla^{\mu}\langle T^{1}_{\mu\nu}\rangle$ turns out to vanish, we can still extract the graviton mass following the result of Equ.~(\ref{eq:coreZ}). We first observe that when $\nabla^{\mu}\langle T^{1}_{\mu\nu}(x)\rangle$ vanishes Equ.~(\ref{eq:coreZ}) becomes the standard St\"{u}ckelberg trick in the discussion of massive gravity theories \cite{Rubakov:2008nh,Hinterbichler:2011tt,deRham:2014zqa} where we introduce an auxiliary field $V^{\mu}(x)$ into the gravitational path integral and shift the graviton field $h^{\mu\nu}$ to
\begin{equation}
h_{\mu\nu}(x)\rightarrow h'_{\mu\nu}(x)= h_{\mu\nu}(x)+\nabla_{\mu}V_{\nu}(x)+\nabla_{\nu}V_{\mu}(x)\,,
\end{equation}
in the action of $h_{\mu\nu}$ (and the matter fields coupled to it) with the resulting action invariant under the gauge symmetry
\begin{equation}
h_{\mu\nu}\rightarrow h_{\mu\nu}(x)+\nabla_{\mu}V_{\nu}(x)+\nabla_{\nu}V_{\mu}(x)\,,\quad V_{\mu}(x)\rightarrow V_{\mu}(x)-\epsilon_{\nu}(x)\,.
\end{equation}
More precisely, in our case, when $\nabla^{\mu}\langle T^{1}_{\mu\nu}(x)\rangle=0$, we would have
\begin{equation}
\begin{split}
Z_{\text{full}}=&\int D[\phi]_{g} D[h_{\mu\nu}]D[V^{\rho}]e^{i S[\phi,h;g]}\,\\=&\int D[\phi]_{g}D[h'_{\mu\nu}]D[V^{\rho}]e^{iS[\phi,h';g]}\,\\=&\int D[\phi]_{g}D[h_{\mu\nu}]D[V^{\rho}]e^{i S[\phi,h';g]}\\=&\int D[\phi]_{g}D[h_{\mu\nu}]D[V^{\rho}]e^{iS[\phi,h;g]+2i\sqrt{16\pi G_{N}}\int d^{d+1}x\sqrt{-g}\nabla^{\mu}V^{\nu}(x)T^{1}_{\mu\nu}(x)}\,.\label{eq:resol}
\end{split}
\end{equation}
As a result we can extract the graviton mass following the same calculation as in Sec.~\ref{sec:demon}.

Here one might think that what we did in Equ.~(\ref{eq:resol}) is a trivial operation as the term $\int d^{d+1}x\sqrt{-g}\nabla^{\mu}V^{\nu}(x)T^{1}_{\mu\nu}(x)$ looks to be zero due to the facts that we can integrate by part and that the matter stress-energy tensor is locally divergenceless. Firstly this is only true if the vector field $V^{\mu}(x)$ decays near the asymptotic boundary $z\rightarrow0$ as $\mathcal{O}(z^{2})$. Though the calculation in Sec.~\ref{sec:demon} suggests that this is not always the case as the effective action for $V^{\mu}$ is proportional to
\begin{equation}
\int d^{d+1}x\sqrt{-g}\Big[\nabla_{\mu}V_{\nu}(x)+\nabla_{\nu}V_{\mu}(x)\Big]\Big[\nabla^{\mu}V^{\nu}(x)+\nabla^{\nu}V^{\mu}(x)\Big]\,,
\end{equation}
which gives the equation of motion
\begin{equation}
\begin{split}
&\nabla^{2}V_{\nu}(x)+\nabla_{\mu}\nabla_{\nu}V^{\mu}(x)\,\\&=\nabla^{2}V_{\nu}(x)+R_{\mu\nu}V^{\mu}(x)\,\\&=\nabla^{2}V_{\nu}(x)-3V_{\nu}(x)\,\\&=0\,,
\end{split}
\end{equation}
where in the second step we used the fact that $V^{\mu}(x)$ is divergenceless as we only consider the conformally coupled matter field. The $z$-component equation of motion is
\begin{equation}
(\partial_{z}^{2}+\partial^{a}\partial_{a}-\frac{2}{z}\partial_{z}-\frac{4}{z^2})V_{z}(x)=0\,,
\end{equation}
which is equivalent to a scalar field with mass square $m^{2}=4$. Hence when there is a source for $V_{\mu}(x)$ on the boundary the leading asymptotic behavior is
\begin{equation}
V_{z}(x)\sim z^{d-\Delta}\,,\quad \text{as } z\rightarrow0\,,
\end{equation}
with $\Delta=\frac{d}{2}+\sqrt{\frac{d^2}{4}+m^2}=\frac{3}{2}+\sqrt{\frac{9}{4}+4}=4$. As result we have
\begin{equation}
V^{z}(x)\sim z^{d-\Delta+2}=z\,.
\end{equation}
When there is no boundary source for $V_{\mu}(x)$ the leading asymptotic behavior is
\begin{equation}
V_{z}(x)\sim z^{\Delta}=z^{4}\,,\quad \text{as } z\rightarrow0\,,
\end{equation}
and
\begin{equation}
V^{z}(x)\sim z^{\Delta+2}=z^{6}\,.
\end{equation}
This suggests that $\int d^{d+1}x\sqrt{-g}\nabla^{\mu}V^{\nu}(x)T^{1}_{\mu\nu}(x)$ is not trivially zero if $V_{\mu}(x)$ has a boundary source. Secondly, it is not always true that the local divergenceless of the matter stress-energy tensor can be promoted to an exact operator equation. The reason is that when the corresponding symmetry i.e. diffeomorphism is spontaneously broken there can be contact terms in the divergence of the two-point function of the matter stress-energy tensor. As we have seen in Sec.~\ref{sec:demon} the coefficient of this contact term is the graviton mass. As a result, we can see that what we did in Equ.~(\ref{eq:resol}) is not always a trivial operation.

\bibliographystyle{JHEP}
\bibliography{main}
\end{document}